\documentclass[fleqn,usenatbib]{mnras}
\usepackage{newtxtext,newtxmath}
\usepackage[T1]{fontenc}
\DeclareRobustCommand{\VAN}[3]{#2}
\let\VANthebibliography\thebibliography
\def\thebibliography{\DeclareRobustCommand{\VAN}[3]{##3}\VANthebibliography}
\usepackage{graphicx}	
\usepackage{amsmath}	
\usepackage{float}
\usepackage{caption}
\usepackage{subcaption}
\usepackage{dblfloatfix}
\usepackage{mathtools}

\newcommand{\ttt}{\texttt}
\newcommand{\be}{\begin{equation}}
\newcommand{\ee}{\end{equation}}
\newcommand{\hmpcc}{\,h^{3}\,{\rm Mpc}^{-3}}
\newcommand{\hmpc}{\,h^{-1}\,{\rm Mpc}}
\newcommand{\hmsun}{\,h^{-1}\,{\rm M}_\odot}

\title[Predicting galaxy assembly bias with ML]{Predicting halo occupation and galaxy assembly bias with machine learning}
\author[Xu et al.]{
Xiaoju Xu,$^{1}$\thanks{E-mail: xiaoju.xu@case.edu}
Saurabh Kumar,$^{1}$\thanks{E-mail: saurabh.kumar@case.edu} 
Idit Zehavi$^{1}$
and Sergio Contreras$^{2}$
\\
$^{1}$Department of Physics Case Western Reserve University, 10900 Euclid Avenue, Cleveland, OH 44106, USA\\
$^{2}$Donostia International Physics Center (DIPC), Manuel Lardizabal Ibilbidea, 4, 20018 Donostia, Gipuzkoa, Spain\\
}
\date{Accepted XXX. Received YYY; in original form ZZZ}

\pubyear{2021}

\begin{document}
\label{firstpage}
\pagerange{\pageref{firstpage}--\pageref{lastpage}}
\maketitle

\begin{abstract}

Understanding the impact of halo properties beyond halo mass on the clustering of galaxies (namely galaxy assembly bias) remains a challenge for contemporary models of galaxy clustering. We explore the use of machine learning to predict the halo occupations and recover galaxy clustering and assembly bias in a semi-analytic galaxy formation model.  For stellar-mass selected samples, we train a Random Forest algorithm on the number of central and satellite galaxies in each dark matter halo. With the predicted occupations, we create mock galaxy catalogues and measure the clustering and assembly bias. Using a range of halo and environment properties, we find that the machine learning predictions of the occupancy variations with secondary properties, galaxy clustering and assembly bias are all in excellent agreement with those of our target galaxy formation model. Internal halo properties are most important for the central galaxies prediction, while environment plays a critical role for the satellites. Our machine learning models are all provided in a usable format. We demonstrate that machine learning is a powerful tool for modelling the galaxy-halo connection, and can be used to create realistic mock galaxy catalogues which accurately recover the expected occupancy variations, galaxy clustering and galaxy assembly bias, imperative for cosmological analyses of upcoming surveys.

\end{abstract}




\section{Introduction}
The advent of large galaxy surveys has transformed the study of large scale structure, allowing high-precision measurements of galaxy clustering statistics. Imaging and spectroscopic surveys, such as the Sloan Digital Sky Survey (SDSS, \citealt{York2000}), the Dark Energy Survey (DES, \citealt{Abbott2016}), the Dark Energy Spectroscopic Instrument (DESI, \citealt{collaboration2016a}), and the upcoming Legacy Survey of Space and Time (LSST, \citealt{LSST2009,Ivezic2019}), provide extraordinary opportunities to utilize such clustering measurements to study both galaxy formation and cosmology. However, it is difficult to model these directly since they depend on complex baryonic processes that are not fully understood. In the standard framework of $\Lambda$CDM cosmology, galaxies form and evolve in dark matter haloes \citep{White1978}, and therefore galaxy clustering can be modelled through halo clustering and galaxy-halo connection.

The formation and evolution of the dark matter haloes are dominated by gravity and their abundance and clustering can be well predicted by analytic models \citep{Press74,Bond91,Mo96,Sheth99,Paranjape2013} and by using high-resolution cosmological numerical simulations \citep{Springel2005,Prada2012,Villaescusa-Navarro2019,Wang2020}. Numerical $N$-body simulations track the evolution of dark matter particles under the influence of gravity and are able to accurately reproduce non-linear clustering on small scales. Haloes or subhaloes can be identified \citep{Springel2001b,Behroozi2013} and merger tree can then be constructed by linking the haloes or subhaloes to their progenitors and descendants at each snapshot in the simulation.

A useful approach for incorporating the predictions of galaxy formation physics is with semi-analytic modelling (SAM), in which the simulated dark matter haloes are populated with galaxies and evolved according to specified prescriptions for gas cooling, galaxy formation, feedback processes, and merging \citep{DeLucia2007,Guo2011,Guo2013,Croton2016,Stevens2016,Cora2018}.  Such models have been successful in reproducing several measured properties of galaxy populations and have become a popular method to explore the galaxy-halo connection.
An alternative approach to model galaxy formation is provided by cosmological hydrodynamic simulations \citep{Schaye2015,Nelson2019}, which simulate both the dark matter particles and the stellar and gas components. The baryonic processes are tracked by a combination of fluid equations and subgrid prescriptions. Cosmological hydrodynamical simulations are starting to play a major role in studying galaxy formation, but are computationally expensive for the large volumes involved. 

Empirical models such as halo occupation distribution (HOD) modelling \citealt{Berlind2002,Cooray2002,Zheng2005,Zehavi2005,Zehavi2011}) and subhalo abundance matching (SHAM, \citealt{Conroy2006,Behroozi2010,Reddick2013,Guo2016,Chaves-Montero16,Contreras2020}) are also used to model galaxy clustering by characterizing the relation between galaxies and their host haloes. In the HOD approach, one fits or utilizes a model for the halo occupation function, the average number of central and satellite galaxies in the host halo as a function of the halo mass. In contrast, the SHAM methodology connects galaxies to dark matter (sub)haloes using a monotonic relation between the galaxy's luminosity (or stellar mass) and the subhalo mass (or maximum circular velocity). Compared to SAM and hydrodynamic simulations, HOD and SHAM are practical and faster ways to generate realistic galaxy mock catalogues, increasingly important for the planning and analysis of galaxy surveys.

In the standard HOD or SHAM approaches, the galaxy content only depends on the halo or subhalo mass (or related mass indicators). However, halo clustering has been shown to depend on secondary halo properties or more generally on the assembly history or large-scale environment of the haloes, a phenomenon termed (halo) assembly bias \citep{Sheth2004,Gao2005,Wechsler2006,Gao2007,Paranjape2018,Ramakrishnan2019}. The dependences on these secondary parameters manifest themselves in different ways and are not trivially described \citep{Mao2018,Salcedo2018,Xu2018,Han2019}. Halo assembly bias might impact large scale galaxy clustering as well, if the formation of galaxy is correlated to that of the host halo, an effect commonly referred to as galaxy assembly bias (GAB hereafter; e.g., \citealt{Croton2007,Zu2008,Chaves-Montero16,Contreras2019,Xu2020,Xu2021}). In such a case, the halo occupation by galaxies will no longer depend solely on halo mass, but will vary with these secondary halo and environmental properties. These expected occupancy variations have recently been studied in SAM and hydrodynamical simulations \citep{Zehavi2018,Zehavi2019,Artale2018,Bose2019,Xu2021}).  

If the GAB is significant in the real universe, neglecting it would have direct implications for interpreting galaxy clustering and the inferred galaxy-halo connection and cosmological constraints \citep{Zentner2014,McEwen2018,McCarthy2019,Lange2019}. Some extensions to include environment or other halo properties have been suggested (e.g., \citealt{Hearin2016,McEwen2018,Contreras2021,Xu2021}). However, given the complexities involved, it is very hard to develop a scheme which will simultaneously incorporate the occupancy variation (hereafter OV) of all relevant halo properties. Moreover, as demonstrated in \citet{Xu2021}, each halo property on its own contributes only a small fraction of the GAB signal, such that a mix of multiple properties will likely be required. this makes first principles predictions for assembly bias challenging. Alternative approaches to predict galaxy properties based on halo assembly history have been proposed \citep{Moster2018,Behroozi2019}, however, the full galaxy-halo connection could be high-dimensional and non-linear, which is difficult to capture by these models.

Machine learning (ML) provides a potentially powerful approach to study the galaxy-halo connection, inferring intricate relations from the complex multi-dimensional data in order to accurately connect the galaxies to the dark matter haloes. In recent years, ML techniques have become a versatile tool with a range of applications in large-scale structure and cosmology \citep{Aragon-Calvo2019,Berger2019,Lucie-Smith2020,deOliveira2020,Arjona2020,Ntampaka2020}. It is also helpful for processing observational data and performing classification \citep{DeLaCalleja2004,Sanchez2014,Tanaka2018,Cheng2020,Wu2020,Mucesh2021,Zhou2021}. In the context of halo modelling, ML can be implemented to predict galaxy properties based on input halo information \citep{Xu2013,Kamdar2016a,Kamdar2016b, Agarwal2018,Wadekar2020,Lovell2021,Moews2021}, and also applied in the reverse sense, predicting halo properties based on galaxy information \citep{Armitage2019,Calderon2019}. More specifically, \citet{Xu2013} make a first attempt to predict the number of galaxies given the halo's properties that can be utilized to create mock catalogues, matching the large scale correlation function to $5\%-10\%$. \citet{Agarwal2018} predict central galaxy properties based on halo properties and environment and find that the average relations of these properties with halo mass are accurately recovered. In \citet{Kamdar2016a,Kamdar2016b}, several galaxy properties such as gas mass, stellar mass, star formation rate, and colour are predicted based on subhalo information. Recently, \citet{Lovell2021} also present a study reproducing several galaxy properties based on subhalo properties in the EAGLE set of hydrodynamic simulations \citep{Schaye2015}.

In this paper, we aim to train a ML model to learn the relation between halo properties and the occupation numbers of galaxies from a galaxy formation simulation. This invariably includes the complex set of effects related to GAB (such as the preferential occupation of galaxies in early-formed haloes as one example). We utilize here Random Forest (RF) classification and regression, one of the most effective ML models for predictive analytics \citep{Breiman2001}.  RF is an ensemble supervised learning method that works by combining decisions from a sequence of base models (decision trees). We use for this purpose stellar mass selected galaxy samples from the \citet{Guo2011} SAM applied to the Millennium Run Simulation \citep{Springel2005}. The input is the halo catalogue including an exhaustive set of halo properties and environment measures and the output will be the occupation numbers of central and satellite galaxies. The RF model will then be used to create mock galaxy catalogues and compared to the true levels of galaxy clustering and large-scale GAB.

 We begin with a RF model that uses all internal and environmental halo properties as input and find an excellent agreement between the predicted HOD, galaxy clustering, and GAB and those measured in the SAM. The RF also provides feature importance which enables us to select the top properties for predicting occupations. Interestingly, the environment properties are found to be important for the satellites occupation but not for central one. We find that using only the top four input features can still recover the full level of GAB. We perform additional tests where we build RF models based on only mass and environment, and alternatively, using the internal halo properties alone.

This methodology can be applied to other galaxy formation models as well, and serve as the basis for an efficient way to populate galaxies in dark matter only simulations, capturing the pertinent information of the galaxy-halo relation and recovering the right level of galaxy clustering including the detailed effects of assembly bias.  Additionally, evaluating the relative feature importance can provide valuable insight regarding the contributors to assembly bias and the importance of halo and environmental properties to galaxy formation and evolution.
Compared to other related ML works which predict the stellar mass of central galaxies (e.g., \citealt{Xu2013,Wadekar2020}; C.\ Cuesta, in prep.), our work utilizes the occupation numbers, more directly probing assembly bias, and allows to naturally incorporate both central and satellite galaxies.
In contrast to \citet{Xu2021} which evaluated the individual contributions to GAB and produced mock catalogues that recover the full level of GAB and OV with respect to specific environment measures,  here we use the full ensemble of properties and are able to reproduce the OV with multiple properties simultaneously. This latter property allows for more realistic and complete mock catalogues, which may be important for certain cosmological applications.

The paper is organized as follows. In Section~\ref{simu}, we briefly describe the $N$-body simulation, the halo and environmental properties, and the SAM galaxy formation model. Section~\ref{ml_measure} provides an introduction to the RF algorithm and the performance measures used to evaluate our models.
In Section~\ref{predictions}, we present our results for the halo occupation, galaxy clustering, and GAB with different combinations of halo and environmental properties. We conclude in Section~\ref{summary}. Appendices~A and B present further results of our analysis.

\section{Dark matter halo and galaxy samples}
\label{simu}

\subsection{$N$-body simulation and halo properties}
\label{simu_halo}

We use in this work the dark matter halo sample from the Millennium $N$-body simulation \citep{Springel2005}. The simulation was run using the GADGET-2 code \citep{Springel2001a}, and adopts the first-year WMAP $\Lambda$CDM cosmology \citep{Spergel2003}, corresponding to the following cosmological parameters: $\Omega_{\rm m}=0.25$, $\Omega_{\rm b}=0.045$, $h=0.73$, $\sigma_8=0.9$, and $n_s$=1. The simulation is in a periodic box with a length of 500 $h^{-1}{\rm Mpc}$ on a side, with $2160^3$ total number of dark matter particles of mass $8.6\times10^8 \hmsun$. The simulation outputs 64 snapshots spanning $z=127$ to $z=0$. At each redshift, the distinct haloes are identified by a friends-of-friends (FoF) group finding algorithm \citep{Davis1985}, and the subhaloes are identified by the \texttt{SUBFIND} algorithm \citep{Springel2001b}. Finally, a halo merger tree is constructed by linking each subhalo to its progenitor and descendant \citep{Springel2005}.

We utilize a set of internal halo properties as well as environmental measures, similar to those used in \citet{Xu2021}, as the input features for the RF models. These halo properties characterise halo structure and assembly history, and the environmental ones measure the density and tidal field at the position of the halo. We list and define all properties used in Table~\ref{table:haloprops}. The halo properties are separated into two categories. The first one are properties that can be obtained from the information from a single snapshot, here the one corresponding to $z=0$, such as $M_{\rm vir}$, $V_{\rm max}$, halo concentration $c$ defined as $V_{\rm max}/V_{\rm vir}$, and specific angular momentum $j$. The second category of halo properties pertains to the assembly history of the haloes and can be calculated from the merger tree. These include $V_{\rm peak}$, $a_{\rm 0.5}$, $a_{\rm 0.8}$, $a_{\rm vpeak}$, the mass accretion rate $\dot M$, $\dot M /M$, $z_{\rm first}$, $z_{\rm last}$, and $N_{\rm merge}$. The environmental properties we use are the mass densities on different smoothing scales, $\delta_{\rm 1.25}$, $\delta_{\rm 2.5}$, $\delta_{\rm 5}$, $\delta_{\rm 10}$, and the tidal anisotropy $\alpha_{1,5}$ \citep{Xu2021}.

\begin{table*}
  \caption{Halo properties and environmental measures used as input features for the RF models. The top part correspond to properties obtained directly from the $z=0$ snapshot in the Millennium database. The middle part are properties computed using the merger tree of the simulation, and the bottom part corresponds to the environmental properties.}
 \centering
 \begin{tabular}{p{0.1\textwidth}p{0.75\textwidth}}
 \hline
 Properties & Definition\\ 
 \hline\hline
 $M_{\rm vir}$ & Halo mass enclosed by the virial radius, defined by 200 times the critical density \\ \hline
 $V_{\rm max}$ & Maximum circular velocity of particles in the halo \\ \hline
 $c$ & Halo concentration, defined as $V_{\rm max}/V_{\rm vir}$ \\ \hline
 $j$ & Specific angular momentum, the angular momentum of the halo normalized by halo mass \\ \hline \hline
 
 $V_{\rm peak}$ & Peak circular velocity, the peak value of maximum circular velocity in the history of the halo \\ \hline
 $a_{\rm 0.5}$ & Scale factor when the halo first reaches 0.5 of its final mass, often referred as the halo formation time (age)\\ \hline
 $a_{\rm 0.8}$ & Scale factor when the halo first reaches 0.8 of its final mass \\ \hline
 $a_{\rm vpeak}$ & Scale factor corresponding to the peak circular velocity  \\ \hline
 $\dot M$ & Halo mass accretion rate \\ \hline 
 $\dot M /M$ & Specific mass accretion rate \\ \hline
 $z_{\rm first}$ & Redshift of the first major merger, defined by a 1:3 mass ratio \\ \hline
 $z_{\rm last}$ & Redshift of the last major merger \\ \hline
 $N_{\rm merge}$ & Total number of the major mergers in the main branch of the merger tree \\ \hline \hline
 
 $\delta_{\rm 1.25}$ & Matter density field at the halo position with a Gaussian smoothing scale of 1.25 $h^{-1}{\rm Mpc}$ \\ \hline
 $\delta_{\rm 2.5}$ & Matter density field at the halo position with a Gaussian smoothing scale of 2.5 $h^{-1}{\rm Mpc}$ \\ \hline
 $\delta_{\rm 5}$ & Matter density field at the halo position with a Gaussian smoothing scale of 5 $h^{-1}{\rm Mpc}$ \\ \hline
 $\delta_{\rm 10}$ & Matter density field at the halo position with a Gaussian smoothing scale of 10 $h^{-1}{\rm Mpc}$ \\ \hline

 $\alpha_{1,5}$ & Tidal anisotropy parameter, defined as $\sqrt{q^2_R}/(1+\delta_{\rm 5})$ where $q^2_R$ is the tidal torque \citep{Paranjape2018}, measured with a $5 \hmpc$ smoothing scale 
  \\ [0.5ex]
\hline
\end{tabular}
\label{table:haloprops}
\end{table*}

\subsection{Galaxy formation model}
\label{sam}

We use the galaxy sample corresponding to the \citet{Guo2011} galaxy formation SAM implemented on the Millennium simulation. It models the main physical processes involved in galaxy formation in a cosmological context.  These processes include reionization, gas cooling, star formation, angular momentum evolution, black hole growth, galaxy merger and disruption, and AGN and supernova feedback. The \citep{Guo2011} is a version of L-galaxies, the SAM code of the Munich group\citep{DeLucia2004,Croton2006,Guo2013,Henriques2015,Henriques2020}, and uses the subhalo merger tree of the simulation to trace and evolve the galaxies through cosmic time. The prescription parameters in the model are tuned to luminosity, colour, abundance, and clustering of observed galaxies. The \citet{Guo2011} SAM model is widely used in literature (e.g., \citealt{Wang2013,Lu2015,Lin2016,Zehavi2018,Xu2021}), and it is publicly available at the
Millennium database \footnote{\url{http://gavo.mpa-garching.mpg.de/Millennium/}}.

When constructing our galaxy samples, we first apply a halo mass cut of $10^{10.7} \hmsun$, below which the number of dark matter particles is too low to reliably host galaxies. We define stellar mass selected samples with different number densities. For our main analysis we focus on a sample with a stellar-mass threshold of $1.42\times10^{10} \hmsun$, corresponding to a number density of $n=0.01 \hmpcc$. This sample includes a total of $745,027$ central galaxies and $505,784$ satellite galaxies. For some of our analysis, we use two additional samples with stellar-mass thresholds of $3.88\times10^{10} \hmsun$ and $0.185\times10^{10} \hmsun$, corresponding to $n=0.00316 \hmpcc$ and $n=0.0316 \hmpcc$, respectively. These three samples are approximately evenly spaced in logarithmic number density and follow the choices made in \citet{Zehavi2018} and \citet{Xu2021}. While the results presented in this paper are limited to the \citet{Guo2011} SAM at z=0, the developed methodology can be applied to any SAM sample and redshift.

\section{Machine learning methodology}
\label{ml_measure}
\subsection {Random forest classification and regression}
\label{MLmethod}

We first briefly discuss the choice of the machine learning model. Linear regression and classification models are the simplest ML models to learn the relation between the input features and the output. However, linear models are limited since even the simplest non-linear transformation (e.g., a polynomial) can lead to a large increase in the number of features, thereby slowing down the learning process. Support vector machines (SVM) are powerful ML algorithms which can transform the input features into higher dimensions without explicitly transforming the features \citep{Aizerman1964,Boser1992}. However, they suffer from increased training time complexity with the size of training data. In contrast, ensemble methods such as Random Forest \citep{Breiman2001} are suitable for our purpose of learning the relation between halo properties and halo occupation because of their ability of dealing with large and high-dimensional datasets.

The Random Forest algorithm combines the output of multiple randomly created Decision Trees to generate the final output. It uses bootstrap aggregation to create random subsets of the training data with replacement on which the decision trees are trained. The decision tree is a flow-like structure in which each internal node represents a ``test'' of an attribute, each branch represents the outcome, and each terminal node or leaf represents the output (the decision taken after computing all attributes). Combining a large number of decision trees, the prediction of RF is the class that is predicted by the majority of the decision trees in the case of RF classification. For RF regression, the prediction is the average prediction from all decision trees.  Thus, for our purpose here, training the RF on a subset of the Millennium halo catalogues and the corresponding SAM galaxy occupations, allows to take into account all the halo properties and predict whether a given halo has a central galaxy or not (classification) and the expected number of satellite galaxies (regression).
 
The main advantage of decision trees is that they perform well with non-linear problems and are computationally cheap since the decision trees can be trained in parallel. One of the major concerns about decision trees is that they can be unstable due to the hierarchical nature of trees: a small change in the training set can result in a difference in the root split which is propagated down to subsequent splits. However, this is mitigated in RF by averaging the predictions over many uncorrelated trees. Decision trees also tend to be strong learners, meaning that individual trees tend to overfit the data. Overfitting is addressed by aggregating the results over many high-variance and low-bias trees. Another important feature of the RF algorithm is that it provides the relative feature importance, i.e the contribution of each input property in making the predictions which we will examine in Section~\ref{predictions}. For a more rigorous discussion of the RF algorithm, we refer the reader to Chapters 9 and 15 of \citet{hastie01} and Chapters 6 and 7 of \citet{Geronml}.

\subsection{Performance measures}
\label{scores}

The RF model includes several `hyper-parameters' which characterize the ensemble of decision trees. In this work we focus on three of them, the total number of the trees in RF, the maximum depth of each tree, and the minimum number of samples in the leaf node of the tree. As common in machine learning analyses, we optimize the performance of the RF algorithm by doing a grid search over these parameters and finding the best fit values. The grid search is performed over 80\% of the full halo catalogue in the simulation, using the so-called 4-fold cross-validation technique (see, e.g., Chapter 5 of \citealt{CVref}). For each choice of hyper parameters, this data is split into four subsets; three are used for training and the remaining one is used for validation and obtaining the ``performance scores''. This is repeated four times so that each of the four subsets is used for validation, and the performance scores are averaged. This process is repeated for each choice on the hyper parameters grid, resulting in the grid point with the highest score.

\begin{figure}
\centering
	\includegraphics[width=0.7\hsize, height =0.7\hsize]{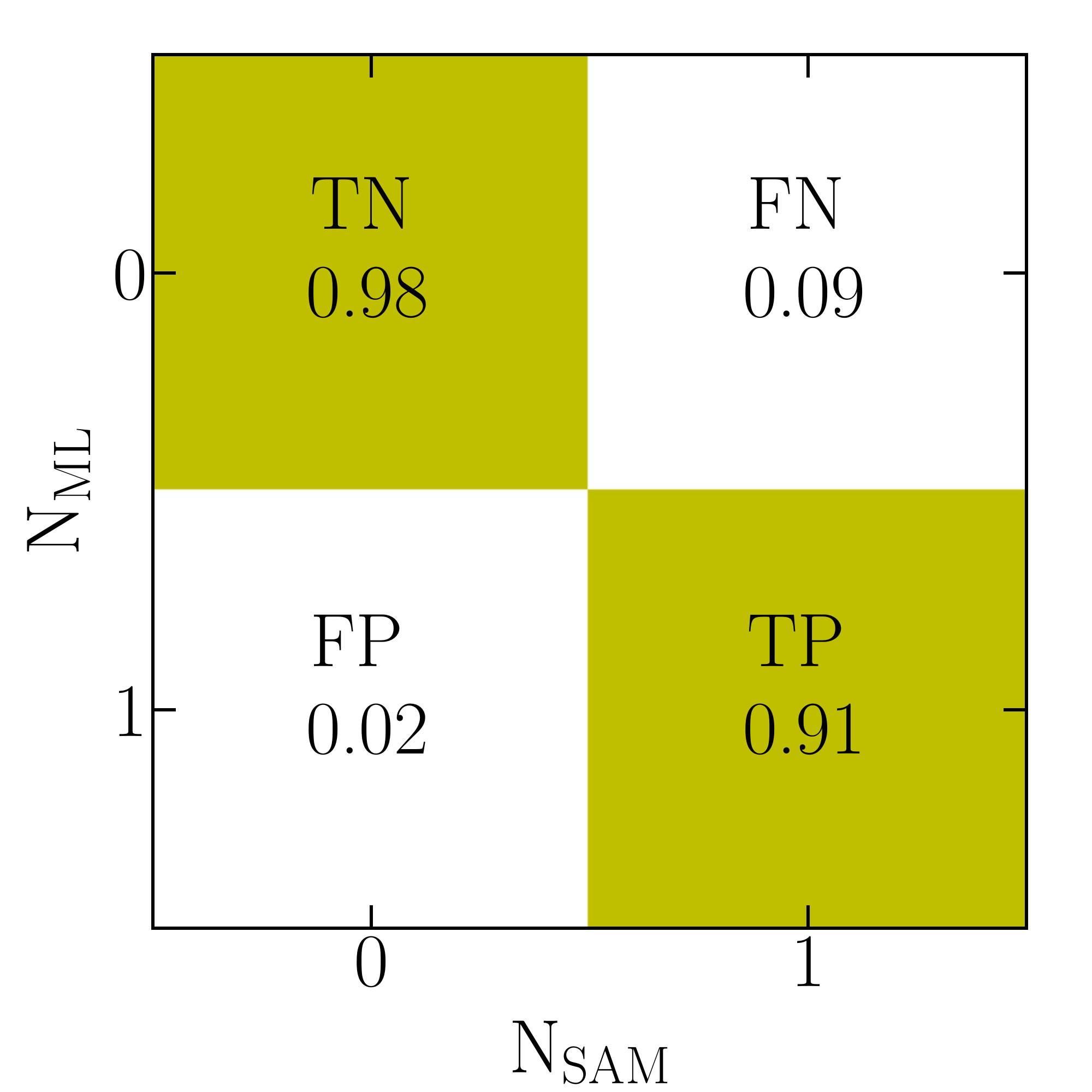}
  \caption{Confusion matrix for central galaxy predictions for the $n=0.01 \hmpcc$ galaxy sample, with all the halo internal and environmental properties used as input. The predictions are obtained from the full sample, with the rows corresponding to the ML predicted values and the columns showing the values in the SAM (see text). 
  }
\label{confusion_mat}
\end{figure}

For classification, a useful way to evaluate its performance is to look at the confusion matrix. To illustrate this we show in Figure~\ref{confusion_mat} the confusion matrix trained using the $n=0.01 \hmpcc$ galaxy sample, using all halo and environmental features. Each row represents the RF predicted class (0 or 1), whereas each column represents the true class in the SAM (0 or 1). In our case, 1 refers to haloes containing a central galaxy and 0 otherwise. Haloes containing central galaxies and predicted as such are referred to as true positives (TP) whereas those predicted as 0 are referred to as false negatives (FN). Haloes without a central galaxy and predicted as such are referred to as true negatives (TN) while those predicted as 1 are false positives (FP). A perfect classifier would have only TN and TP and zero off-diagonal values. The confusion matrix shows the fraction of haloes in each category. We see that, in our case, the fractions of TP and FN are 0.91 and 0.09, respectively, where the predictions are normalized by the total number of haloes containing a central galaxy. The fractions of TN and FN are 0.98 and 0.02, respectively, normalized in this case by the total number of haloes not containing a central galaxy. 

A more concise metric utilizing the confusion matrix is the $F_1$ score defined as:
\be
F_1 = 2PR/[P + R], 
\ee
where $P$ and $R$ are the Precision and Recall. Precision measures the accuracy rate,
\be
P = \text{TP}/[\text{TP} + \text{FP}],
\ee
while the recall, also known as sensitivity or true positive rate, is
\be
R = \text{TP}/[\text{TP} + \text{FN}]. 
\ee   
Since precision and recall measure different aspects of the success of the predictions, they are usually combined to evaluate a classifier.  We use the $F_1$ score, conveying the balance of precision and recall, to optimize the choice of hyper parameters for the RF classification of central galaxies.

For regression, we use the $R^2$ score or the coefficient of determination defined as:
\be
R^2 = 1 - S_{\text{res}}/S_{\text{tot}},
\ee 
where $S_{\text{res}}$ is the residual sum of squares,
\be
S_{\text{res}} = \sum_i (p_i - y_i)^2 ,
\ee
where $p_i$ is the prediction for each input data and $y_i$ the true value. This sum is normalized by the underlying total sum of squares relative to the mean $\bar{y}$:
\be
S_{\text{tot}} = \sum_i (y_i - \bar{y})^2.
\ee
Even though we explored other performance measures, we chose the $R^2$ score to set the hyper parameters for the RF regression predictions of the number of satellite galaxies for the cases we explore. 

We utilize the Python package \ttt{sklearn} for performing all grid searches and RF training. We use 80\% of the full halo catalogue in the Millennium simulation as the training set.  For each application, we first set the RF hyper parameters to those that give the highest scores in the grid search. We then proceed to train the RF classification and regression models to predict the number of central and satellite galaxies in each halo. In practice, when estimating the clustering and GAB, we average the predictions of 10 training sets (each containing 80\% of the total haloes) drawn randomly out of 90\% of the full catalogue. This allows to reduce the sensitivity to the specifics of the training set (though the sets clearly still have a large overlap). The remainder 10\% of the haloes are left as an independent test set, not used for either the training or cross-validation.

\section{Machine learning results}
\label{predictions}

In this section, we present the results of our RF models. For the main analysis described here, we use the stellar-mass selected $n=0.01 \hmpcc$ sample as mentioned in Section~\ref{sam}. The direct predictions output of the ML model are the numbers of central and satellite galaxies in each halo. We comprehensively compare them with the 'true' distribution of the SAM galaxy sample in multiple ways. We first directly compare the galaxy numbers on a halo-by-halo basis. We then compare the halo occupation functions, namely the average number of galaxies as a function of halo mass, as well as the variations in these halo occupation functions with secondary properties (referred here as the OV; e.g., \citealt{Zehavi2018}). We then proceed to populate the halo sample with the predicted number of galaxies to create a mock galaxy catalogue based on the ML predictions. We calculate the clustering of the ML galaxy sample and compare to that of the SAM sample. Finally, we examine and compare the impact of GAB on the large-scale clustering signal. We describe all these in detail below.
We show the results using the full halo catalogue of the Millennium simulation, which includes the training sets, used to build the ML model, and the smaller (10\% of the haloes) test sample. We have repeated our main analysis using only the test sample, finding similar results to the ones shown here.

\subsection{All features}
\label{allfeatures}

Here we present the ML results when using all available features, namely all the internal halo properties and environmental measures specified in Table~\ref{table:haloprops}.  
The accuracy of the ML predictions for hosting a central galaxy with stellar mass larger than our sample's threshold in the individual haloes has already been presented in Figure~\ref{confusion_mat}. Again, we find that for haloes which host a central galaxy above the stellar-mass threshold in the SAM, 91\% of them are predicted to host a central galaxy by our ML model. For haloes that do not host a central galaxy, 98\% of them are accurately predicted as such in our model. The difference in the relative values likely simply reflects the larger number of haloes with no central galaxy for this stellar-mass threshold, such that the number of misclassified haloes is roughly comparable. Note that we do not expect the ML algorithm to provide an accurate prediction for every single halo, due to the stochasticity involved, for example in the scatter between stellar mass and halo mass (and such a case would indicate extreme overfitting in the least). We view this agreement as very good. 

\begin{figure}
	\centering
	\begin{subfigure}[h]{0.48\textwidth}
		\includegraphics[width=\textwidth]{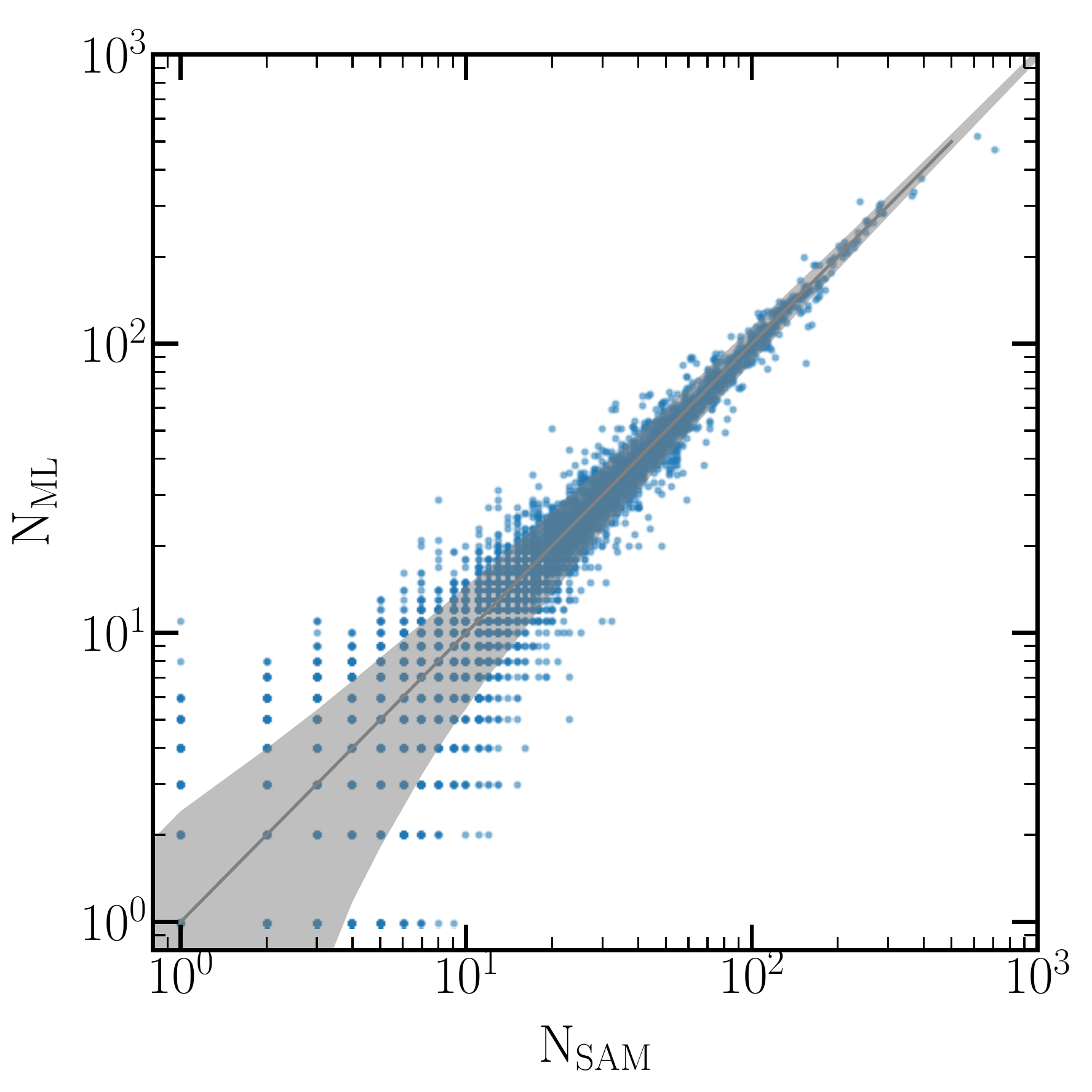}
	\end{subfigure}
	\hfill
\caption{Comparison between the RF predicted number of satellite in each halo and the actual number from the SAM. The blue dots show these values for each individual halo, for the ML model applied to the $n=0.01 \hmpcc$ galaxy sample, using all halo features. The diagonal grey line indicates the idealized case where the number is identical, and the shaded region represents the Poisson error often assumed in HOD models.
}
\label{scatter}
\end{figure}

The `raw' predicted numbers of satellite galaxies from the RF regression model are not required to have an integer value a-priori.  We assign it to the nearest integers following a Bernoulli distribution with this mean. In practice, this amounts to assigning, e.g., 4.3 satellites to 3 with a 70\% probability or to 4 with 30\% probability. The relation between these discrete (integer) predictions for the number of satellites and the SAM number of satellites in each halo is presented in Figure~\ref{scatter}. Each point represents the satellite occupation in a single halo, showing the scatter of the RF predictions along the y-axis. The grey shaded area shows, for comparison, a simple Poisson scatter as is often assumed in HOD modelling (the shaded area appears to increase at low numbers, just due to the log scale plotted). The scatter in the ML prediction is larger than the Poisson scatter, due to the more complex model and limitations of the RF regression. This also suggests that we are not overfitting the data here. Though not shown here, for clarity, we also perform a linear fit of the points to examine any bias in the predictions. For a fully unbiased prediction, the slope of the linear fit would be one. However, we find a slope of 0.96 which indicates a slight underprediction. This is likely caused by the lower ML prediction relative to the SAM at the largest occupation numbers (high halo mass). This underprediction is also found in \citet{Xu2013} and is considered a result of the small number of the most massive haloes in the simulation. Since the level of the underprediction is low, it should not impact the results in this paper.

\begin{figure}
	\centering
	\begin{subfigure}[h]{0.48\textwidth}
		\includegraphics[width=\textwidth]{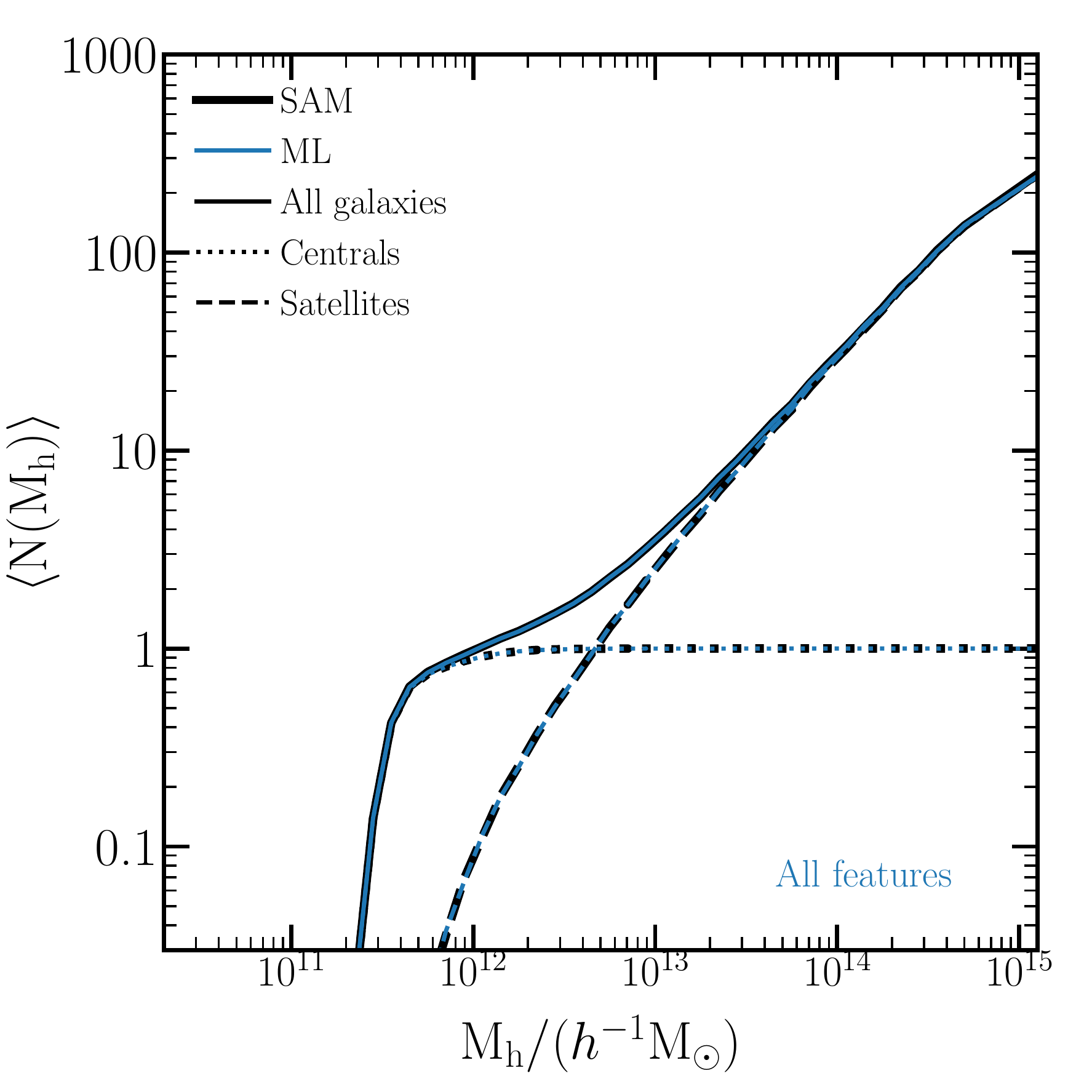}
	\end{subfigure}
	\hfill
	\begin{subfigure}[h]{0.48\textwidth}
		\includegraphics[width=\textwidth]{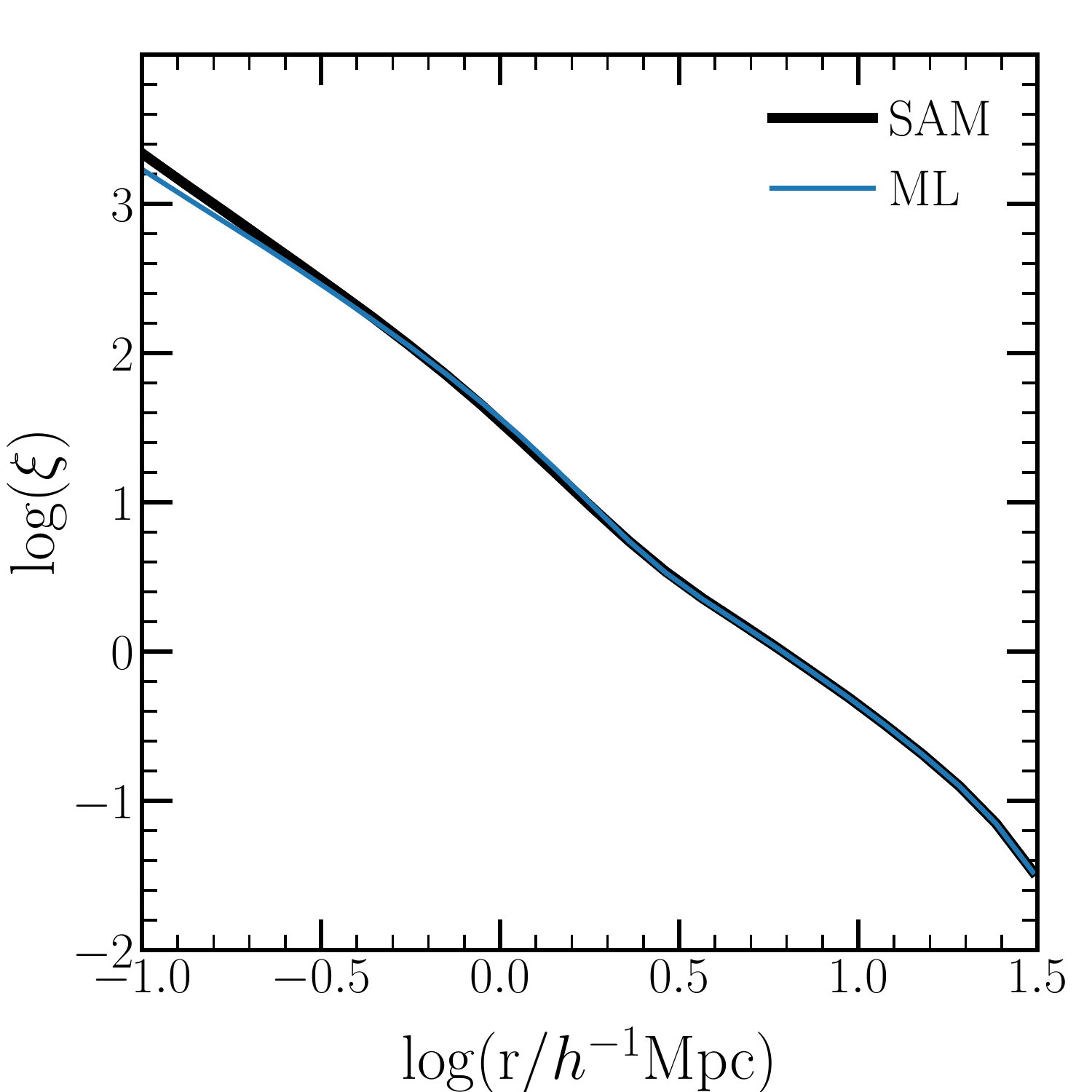}
	\end{subfigure}
	\hfill
        \caption{\textbf{Top}: The halo occupation function for the SAM $n=0.01 \hmpcc$ sample (black) and ML prediction (blue) using all the halo and environmental properties. The individual contributions from central and satellite galaxies are shown as dotted and dashed lines, respectively. \textbf{Bottom}: The galaxy two-point auto-correlation function of the ML prediction (blue) compared to the SAM (black). The small difference on small scales is due to the galaxy profile in the SAM slightly deviating from the NFW profile assumed for the ML prediction.
        }
\label{hodall}
\end{figure}

Moving away from the comparisons on an individual halo basis, we now shift to comparing the central and satellite galaxy numbers averaged in mass bins, namely the halo occupation functions commonly used in the HOD framework. The top panel of Figure~\ref{hodall} compares the halo occupation function corresponding to the ML predictions (blue) with that of the SAM (black) for the $n=0.01 \hmpcc$ galaxy sample. We find that the predictions are in excellent agreement with the halo occupation of the SAM galaxies, as can be seen from the indistinguishable lines.

With the predicted number of central and satellite galaxies in each halo, we populate the haloes and create a mock galaxy catalogue to measure the clustering. For each halo, we place the central galaxy at the halo center and populate satellites with an NFW profile, going out to twice the virial radius. The bottom panel of Figure~\ref{hodall} shows the resulting two-point auto-correlation function relative to that measured from the SAM. Again, we find excellent agreement between the ML predictions and the SAM. On small scales, the prediction deviates from the SAM since an NFW profile is adopted in the mock catalogue, which is slightly different from the radial distribution of the SAM satellites (e.g., \citealt{Jimenez2019}). Since we are focused here on modelling GAB, we will only show our predicted clustering results on large scales (larger than $\sim 7 h^{-1}{\rm Mpc}$) from here on.
%

\begin{figure*}
	\centering
	\begin{subfigure}[h]{0.8\textwidth}
		\includegraphics[width=\textwidth]{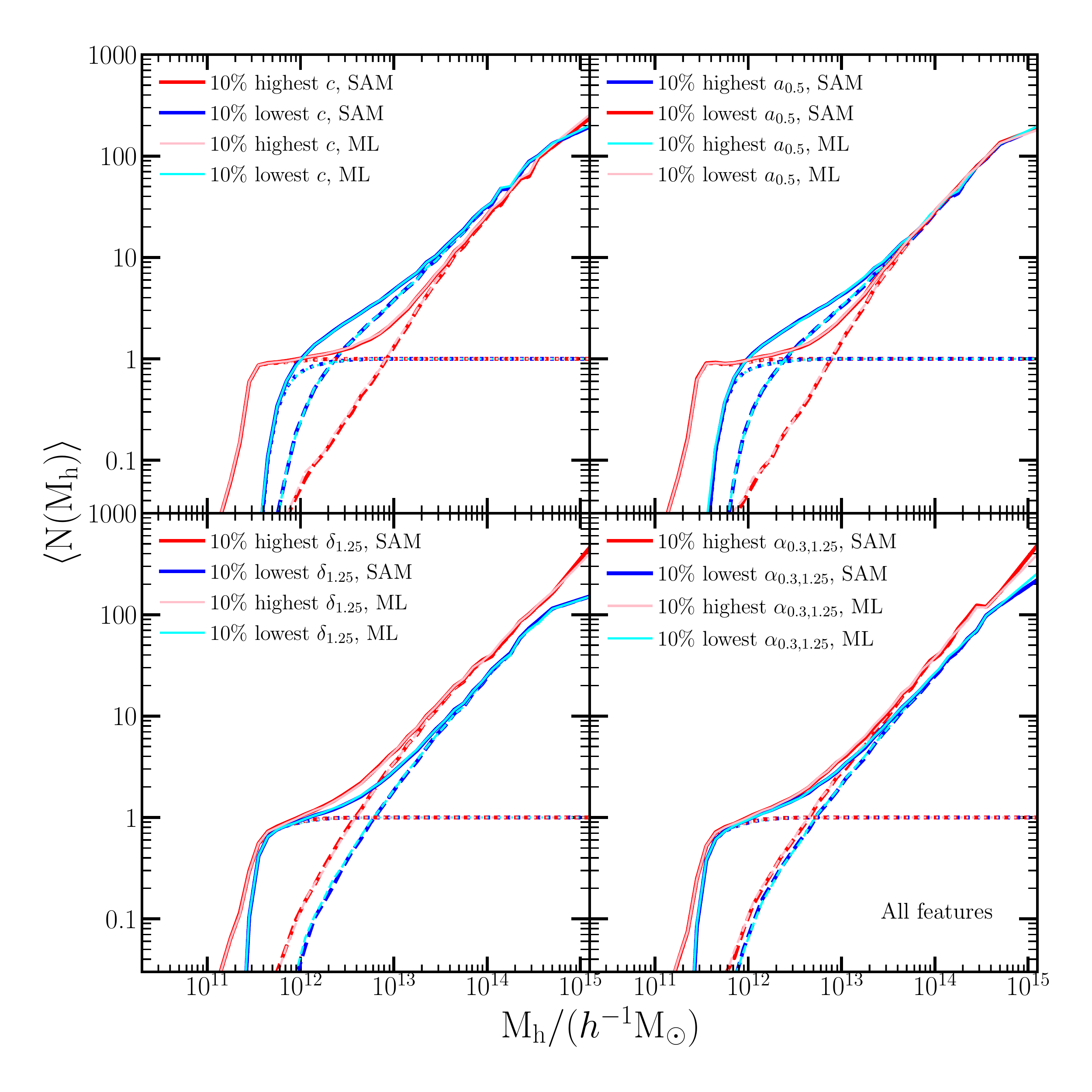}
	\end{subfigure}
	\hfill
\caption{The occupancy variations in the predicted halo occupation functions, when using all halo and environmental properties as input features. Each panel corresponds to a different secondary property, $c$, $a_{\rm 0.5}$, $\delta_{\rm 1.25}$, and $\alpha_{\rm 0.3,1.25}$, as labelled. In all panels, red and blue and lines represent the SAM occupations in the 10\% of haloes with the highest and lowest values, respectively, of the secondary properties in fixed mass bins. Pink and cyan lines show the corresponding cases for the ML predictions. The numbers of centrals, satellites, and all galaxies are shown by dotted, dashed, and solid lines, respectively. 
}
\label{ov_all}
\end{figure*} 

In addition to halo occupation as function of mass, we also examine in detail the variations of the halo occupations with secondary properties. Since halo clustering also depends on such properties (halo assembly bias), together with the OV, galaxy clustering would also be impacted. An HOD model that captures the OV dependence on a specific halo property would thus also capture the GAB caused by this halo property \citep{Xu2021}. These OVs are shown in Figure~\ref{ov_all} for some representative cases of the internal halo properties (concentration, $c$, and halo formation time, $a_{\rm 0.5}$, shown in the top panels) and the environmental measures ($\delta_{\rm 1.25}$ and $\alpha_{\rm 0.3, 1.25}$, shown on the bottom).  Similar to $\alpha_{\rm 1,5}$, $\alpha_{\rm 0.3, 1.25}$ is defined as a measurement of tidal anisotropy on the smoothing scale of $1.25 \hmpc$:
\begin{equation}
\label{alpha03g125}
 \alpha_{\rm 0.3,1.25}=\sqrt{q^2_R}/(1+\delta_{\rm 1.25})^{0.3},
\end{equation}
where $q^2_R$ is the tidal torque measured with the same smoothing scale and the normalization is modified by a 0.3 power \citep{Xu2021}. The red and blue curves in each panel show the occupations for the 10\% of the halo population in each mass bin with the highest and lowest values of the secondary property in the SAM sample, whereas cyan and pink show those predicted by the ML models. Dotted, dashed, and solid curves indicate the central, satellite, and total occupation number. We note that we use $a_{\rm 0.5}$, the scale factor when the halo accretes half of its halo mass, as a proxy for halo age. Highest $a_{\rm 0.5}$ values thus correspond to later formation times and the youngest ages, and vice versa, the earliest formation times correspond to the oldest age (and are colour coded accordingly). 

The OVs shown in Figure~\ref{ov_all} generally follow the trends already examined in detail in previous works \citep{Zehavi2018,Contreras2019,Xu2021}. E.g., older haloes (higher formation time, smaller $a_{\rm 0.5}$ values) tend to start occupying central galaxies at lower halo masses. In contrast, such haloes, host on average less satellites than later-forming haloes.  The striking result in this work is the excellent agreement between the ML predictions and the SAM ones, for all secondary properties. That implies that the RF algorithm is able to accurately learn and reproduce the different secondary trends.
Note that while $\alpha_{\rm 1,5}$ is one of the input features, $\alpha_{\rm 0.3, 1.25}$ is not, and while they may be correlated to some extent, they play different roles in GAB. \citet{Xu2021} show that $\alpha_{\rm 1,5}$ accounts for a small fraction of GAB, whereas $\alpha_{\rm 0.3, 1.25}$ captures the full effect on galaxy clustering.  The tidal anisotropy parameter $\alpha_{\rm 0.3, 1.25}$ is also partially correlated with $\delta_{\rm 1.25}$, but include additional information on the tidal shear. So it is interesting to see that the OV dependence on $\alpha_{\rm 0.3, 1.25}$ can be well reproduced by the ML algorithm, without serving as input for it. More generally, since GAB is a result of halo assembly bias combined with the OV, and the individual OVs are accurately reproduced, we expect that the GAB signal can be well recovered as well.

\begin{figure*}
	\centering
	\begin{subfigure}[h]{0.48\textwidth}
		\includegraphics[width=\textwidth]{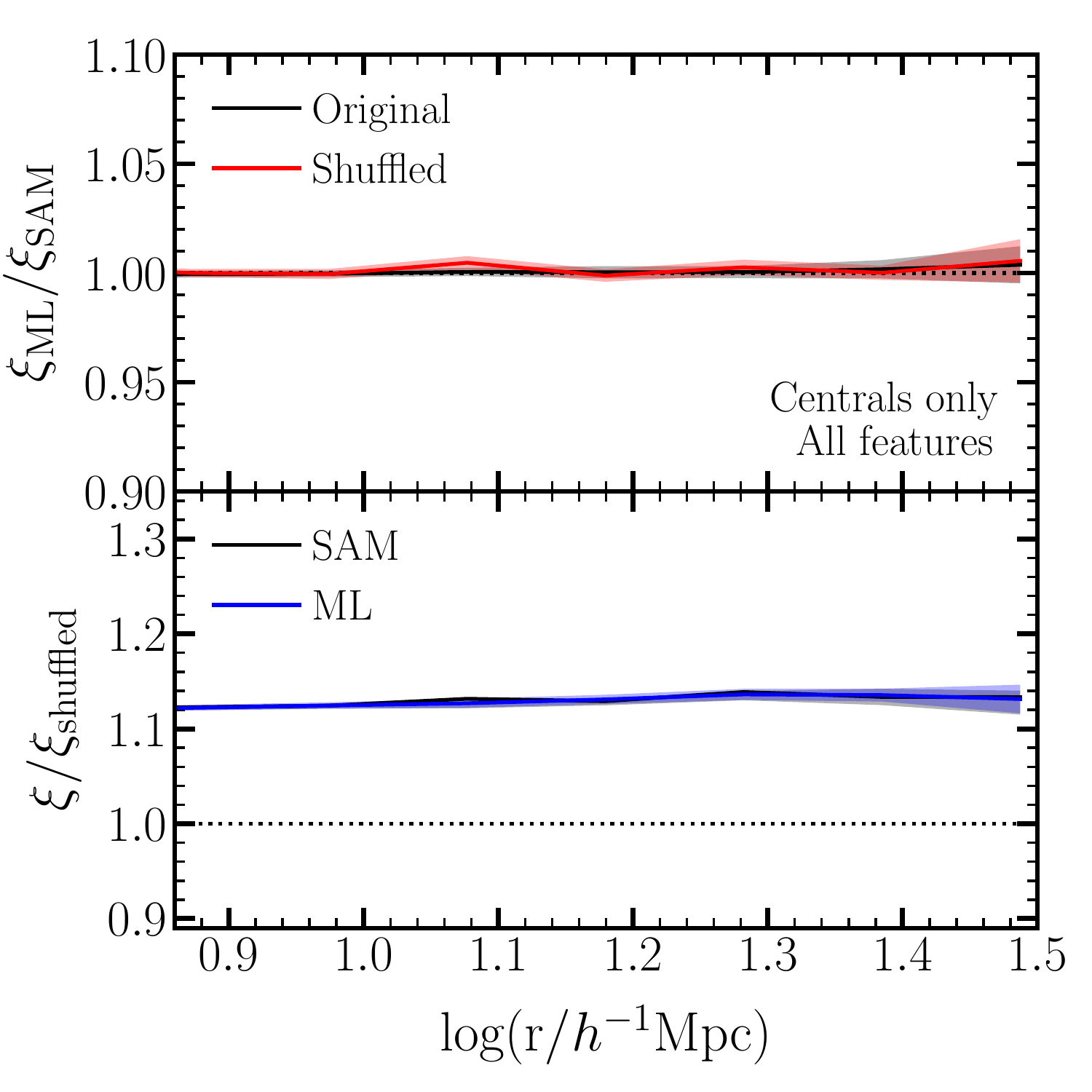}
	\end{subfigure}
	\hfill
	\begin{subfigure}[h]{0.48\textwidth}
		\includegraphics[width=\textwidth]{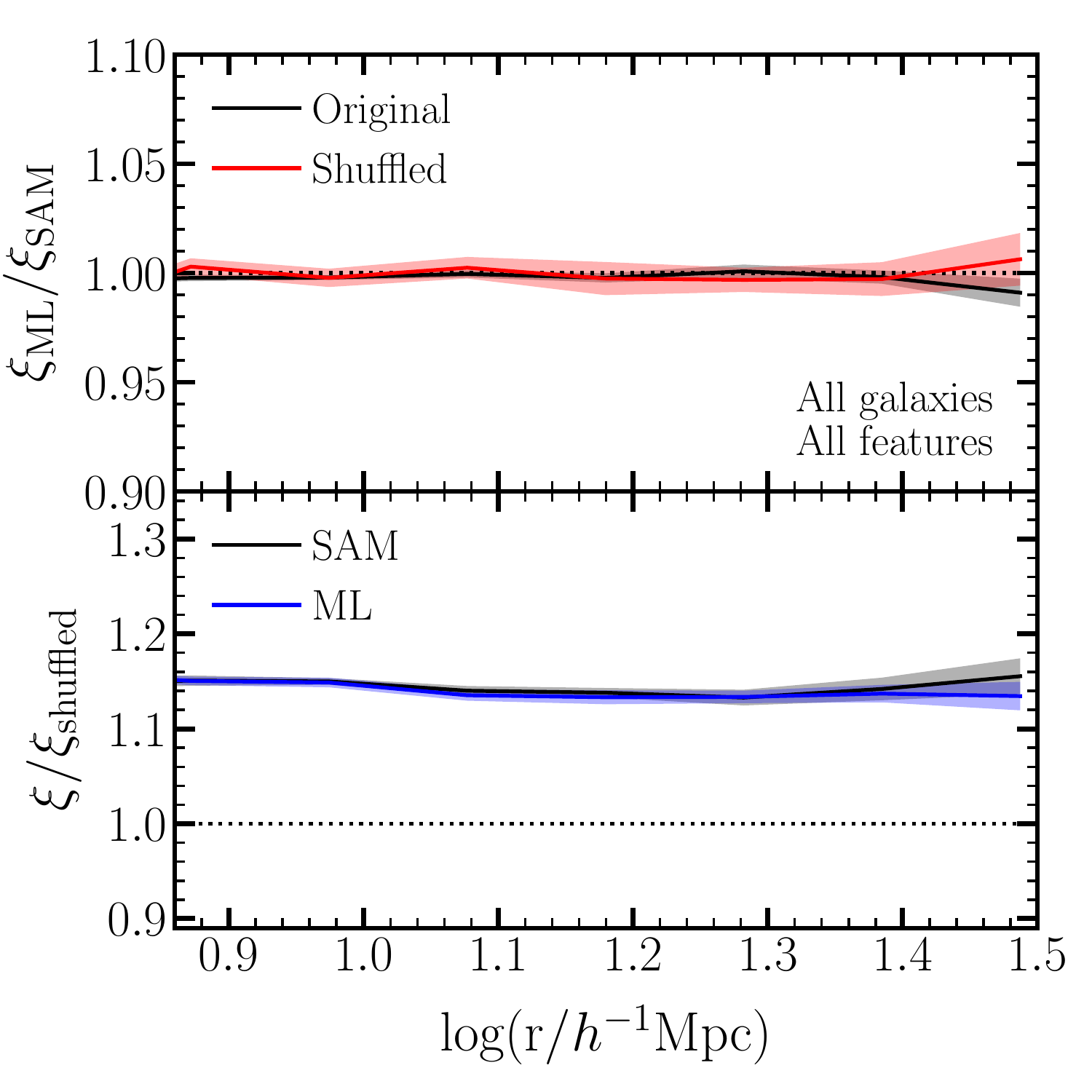}
	\end{subfigure}
	\hfill
        \caption{Comparison of the measured correlation functions and GAB of the SAM and the ML predicted mock catalogue, when using all features. The left-hand side shows these clustering results for central galaxies only, while the right-hand side shows the same for all (central and satellite) galaxies corresponding to the $n=0.01 \hmpcc$ sample.  For both these cases, the top panels show the ratios of the correlation function of the ML-predicted mock catalogues relative to that of the SAM. Ratios of the original (unshuffled) correlation functions are shown in black, while the ratios of the shuffled samples of each are shown in red. The bottom panels (on both sides), show the measured GAB signal, namely the ratio of the original correlation function to that of the shuffled sample.  Here, the SAM GAB measurement is shown in black while the ML GAB is shown in blue. The shaded areas, in all panels, indicate the error bar measured from 10 different shuffled samples of the SAM galaxies and the 10 different realizations of the RF model.
}
\label{cluster_gab_all}
\end{figure*}  

The GAB signature is usually measured as the ratio between the correlation function of the galaxy sample and that of a shuffled sample, created by randomly reassigning the galaxies among haloes of the same mass \citep{Croton2007}. The shuffling process effectively removes the connection of the galaxies to the assembly history of the haloes and eliminates the dependence on any secondary property other than halo mass (i.e it erases all OVs). Comparison between the clustering of the shuffled sample and the original thus reveals the overall effect of GAB, typically seen as an increased clustering amplitude on large scales. Following standard practice \citep{Croton2007,Zehavi2018,Contreras2019,Xu2021}, we shuffle the central galaxies and then move the satellites together with their associated central galaxy. This results in the shuffled sample having the same clustering as the original sample on small (one-halo) scales. 

These results are examined in detail in Figure~\ref{cluster_gab_all}, showing the different large-scale clustering measurements separately for the central galaxies only on the left-hand side and for the full (central and satellite galaxies) sample on the right. We already saw in Figure~\ref{hodall} that the overall clustering of the ML mock sample is highly consistent with that of the SAM on large scales. This is presented more clearly in the top panels of Figure~\ref{cluster_gab_all}, where the black line shows the ratio of the ML predicted clustering to that of the SAM. The shaded regions hereafter indicate the uncertainty associated with the 10 different training sets (see \S~\ref{scores}). In both cases, we see that the SAM clustering is accurately reproduced. 
Our results are a vast improvement compared to \citet{Xu2013} who recover the amplitude of galaxy clustering to 5\%-10\% using the halo occupations as well. We reproduce the clustering to sub-percent precision, perhaps due to both using a larger training sample and including also environmental properties. The latter is in line with recent studies that demonstrate the important role of environment in accurately capturing the level of galaxy clustering \citep{Hadzhiyska2020,Xu2021}. 
We then proceed to examine the results of the shuffled samples. We shuffle each of the SAM sample and the ML mock sample in bins of fixed halo mass, as described above.  The ratios of the shuffled ML predicted clustering to that of the shuffled SAM clustering are presented as the red lines in the top panels of Figure~\ref{cluster_gab_all}. Once again, these ratios are extremely close to unity, indicating an excellent agreement between the shuffled ML clustering and the shuffled SAM clustering.

We examine directly the GAB signature in the bottom panels of Figure~\ref{cluster_gab_all}. Namely, we present ratios of the large-scales correlation function of the original sample to that of the shuffled sample, $\xi/\xi_{\rm shuffled}$. Black lines represent this ratio, i.e the GAB signal, in the SAM while the blue lines represent the ML-predicted GAB signal. The error bar on the SAM measurement is the scatter from 10 different shuffled samples, while the error bar on the ML predictions arises from the 10 different training sets (each with its own shuffled sample). Again, this is shown for the central galaxies only on the left-hand side and for the full samples, including satellites, on the right.
These ratios have already been studied with this specific SAM sample \citep{Zehavi2018,Zehavi2019,Xu2021}. The roughly 15\% increase of clustering in the original SAM sample versus the shuffled one arises from the differentiated occupation of haloes with galaxies according to secondary halo properties which exhibit halo assembly bias. For example, galaxies tend to preferentially occupy older haloes which exhibit stronger clustering, resulting in an increased large-scale galaxy clustering (GAB). We note, again, that the excess clustering shown here is the overall combined effect from all secondary properties.

The remarkable result clearly shown in the bottom panels of Figure~\ref{cluster_gab_all} is the excellent agreement between the GAB signal measured by the ML-predicted sample and that of the original SAM galaxy sample. This is exhibited by the nearly perfect agreement between the blue and black lines in each panel, for central galaxies only (left) and for the full sample (right).  The RF model applied trained on the individual halo occupations is thus able to accurately reproduce the GAB effect in the large-scale galaxy clustering. Together with the recovered OVs, we see that the ML model is highly successful in reproducing all aspects of the complex phenomena of assembly bias.

A simple measure of the agreement between the GAB signals, beyond the striking agreement by eye,  is provided by 
\begin{equation}
\label{eq:fab}
f_{\rm AB}=\langle (\xi_{\rm ML}/\xi_{\rm shuffled,ML} - 1) / (\xi_{\rm SAM}/\xi_{\rm shuffled,SAM} - 1) \rangle,
\end{equation}
which represents the recovered fraction of GAB. The averaging is done over the clustering ratio values measured on large scales of $9 \sim 30 h^{-1}{\rm Mpc}$. For the cases shown in the bottom panels of Figure~\ref{cluster_gab_all}, namely the $n=0.01 \hmpcc$ sample using all the available features in the ML model, we obtain nearly perfect recovery with $f_{\rm AB}=0.99$ for the central galaxies only case and $f_{\rm AB}=0.98$ for the full sample (i.e they recover the full GAB signal to 1-2\%). The recovered level of the correlation function can be similarly estimated as $\langle \xi_{\rm ML}/\xi_{\rm SAM} \rangle$, returning a value of 1.00 for both these cases (to the level of accuracy quoted).
These values are summarized in Table~\ref{table:clustering}, for all the cases explored in this paper, and include also the values of the $F_1$ and $R^2$ performance scores of the RF predictions (\S~\ref{scores}). 
The results of the RF models with all features are listed in the top two lines of Table~\ref{table:clustering}.  The following lines in the table are the results of other RF models with different sets of input features as labelled, for which we provide more details and discussion in the following subsections. Table~\ref{table:clustering} also includes the values obtained using all features for two additional stellar-mass selected galaxy samples corresponding to $n=0.00316 \hmpcc$ and $n=0.0316 \hmpcc$. The clustering and GAB results for these two samples are presented in Appendix~\ref{num_den}.

\begin{table*}
  \caption{Prediction results for RF models with different input features. The first two columns indicate the input features for the central and satellite galaxies. The centrals-only cases are indicated by a ``$-$'' in the second (satellite) column. The performance scores $F_{1}$ and $R^2$ for the centrals and satellites are listed in the third and fourth columns, respectively. The next column shows the recovered fraction of the correlation function, $\langle \xi_{\rm ML}/\xi_{\rm SAM} \rangle$, averaged over scales of $9 \sim 30 h^{-1}{\rm Mpc}$. We do not include a separate column for this property measured for the shuffled samples, since its accuracy is 1.00 (within the significance quoted) for {\it all} cases shown. The final column represents the accuracy of recovering the GAB signal using the $f_{\rm AB}$ measure. The main predictions are all based on the galaxy sample of number density $n=0.01 \hmpcc$ and are listed in top 10 lines. The predictions with all features for two other number densities $n=0.00316 \hmpcc$ and $n=0.0316 \hmpcc$ are listed at the bottom. }
\centering
\setlength{\tabcolsep}{6pt}
 \begin{tabular}{c c c c c c} 
 \hline
 input (cen) & input (sat) & $F_1$ score & $R^2$ score & recovered $\xi$ & recovered $f_{\rm AB}$\\ 
 \hline\hline
 all & -- & 0.89 & -- & 1.00 & 0.99 \\ \hline
 all & all & 0.89 & 0.94 & 1.00 & 0.98 \\ \hline
 top 4 & -- & 0.88 & -- & 1.00 & 0.97 \\ \hline
 top 4 & top 4 & 0.88 & 0.93 & 1.00 & 1.00 \\ \hline 
 $M_{\rm vir}$+$\delta_{\rm 1.25}$ & -- & 0.79 & -- & 0.99 & 0.86 \\ \hline
 $M_{\rm vir}$+$\delta_{\rm 1.25}$ & $M_{\rm vir}$+$\delta_{\rm 1.25}$ & 0.79 & 0.91 & 0.99 & 0.92 \\ \hline
 internal & -- & 0.89 & -- & 1.00 & 0.99 \\ \hline
 internal & internal & 0.89 & 0.91 & 0.97 & 0.70 \\ \hline
 single-epoch & -- & 0.85 & -- & 1.00 & 1.00 \\ \hline
 single-epoch & single-epoch & 0.85 & 0.91 & 0.99 & 0.95 \\ 
 \hline \hline
all (n=0.00316) & -- & 0.74 & -- & 0.98 & 0.83 \\ \hline
all (n=0.00316) & all (n=0.00316) & 0.74 & 0.87 & 0.99 & 0.96 \\ \hline
all (n=0.0316) & -- & 0.96 & -- & 1.00 & 1.00 \\ \hline
all (n=0.0316) & all (n=0.0316) & 0.96 & 0.95 & 1.00 & 0.99 \\ [0.5ex]
\hline
\end{tabular}
\label{table:clustering}
\end{table*}

\subsection{Feature importance }
\label{feature_imp}

\begin{figure*}
	\centering
	\begin{subfigure}[h]{0.46\textwidth}
		\includegraphics[width=\textwidth]{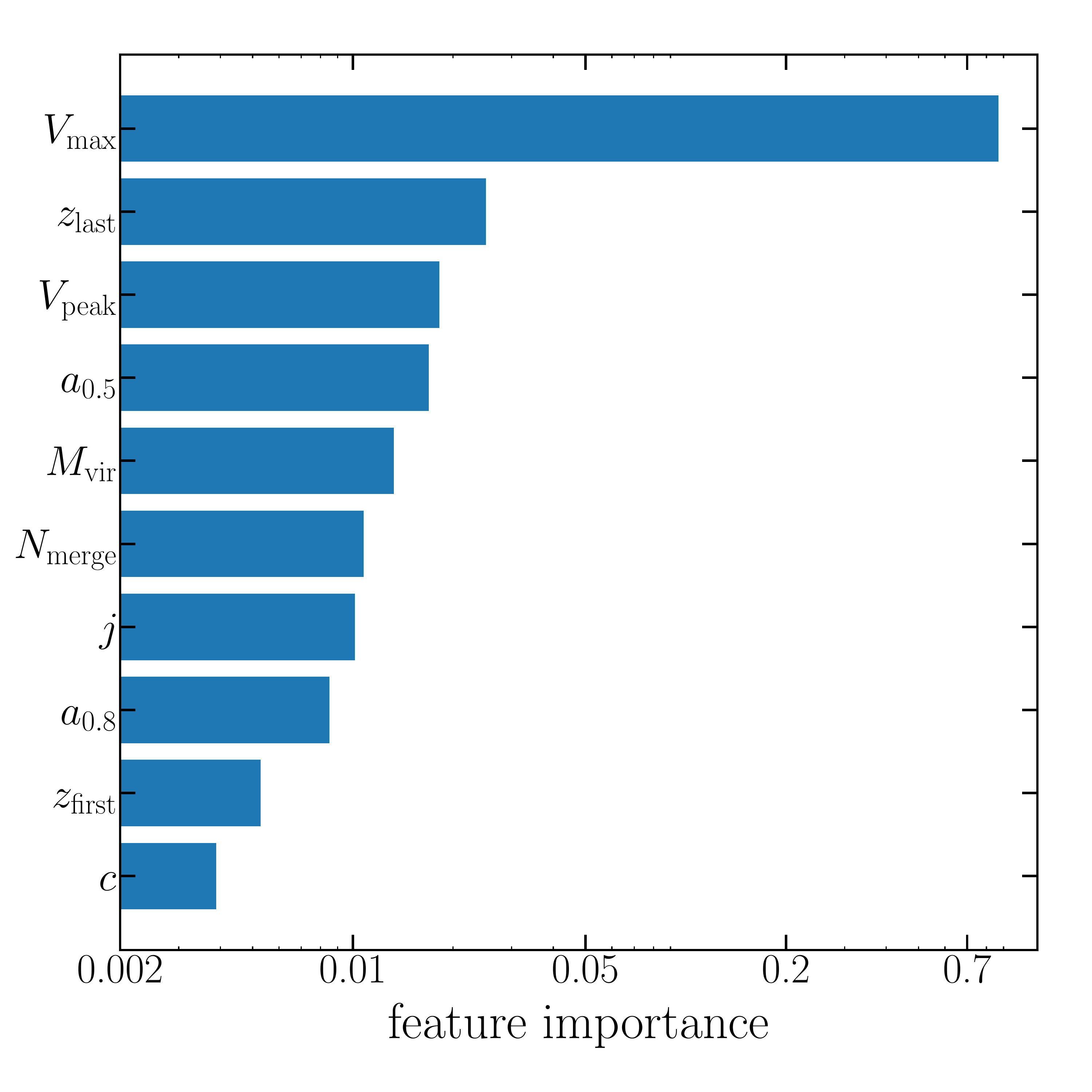}
	\end{subfigure}
	\hfill
	\begin{subfigure}[h]{0.52\textwidth}
	\includegraphics[width=\textwidth]{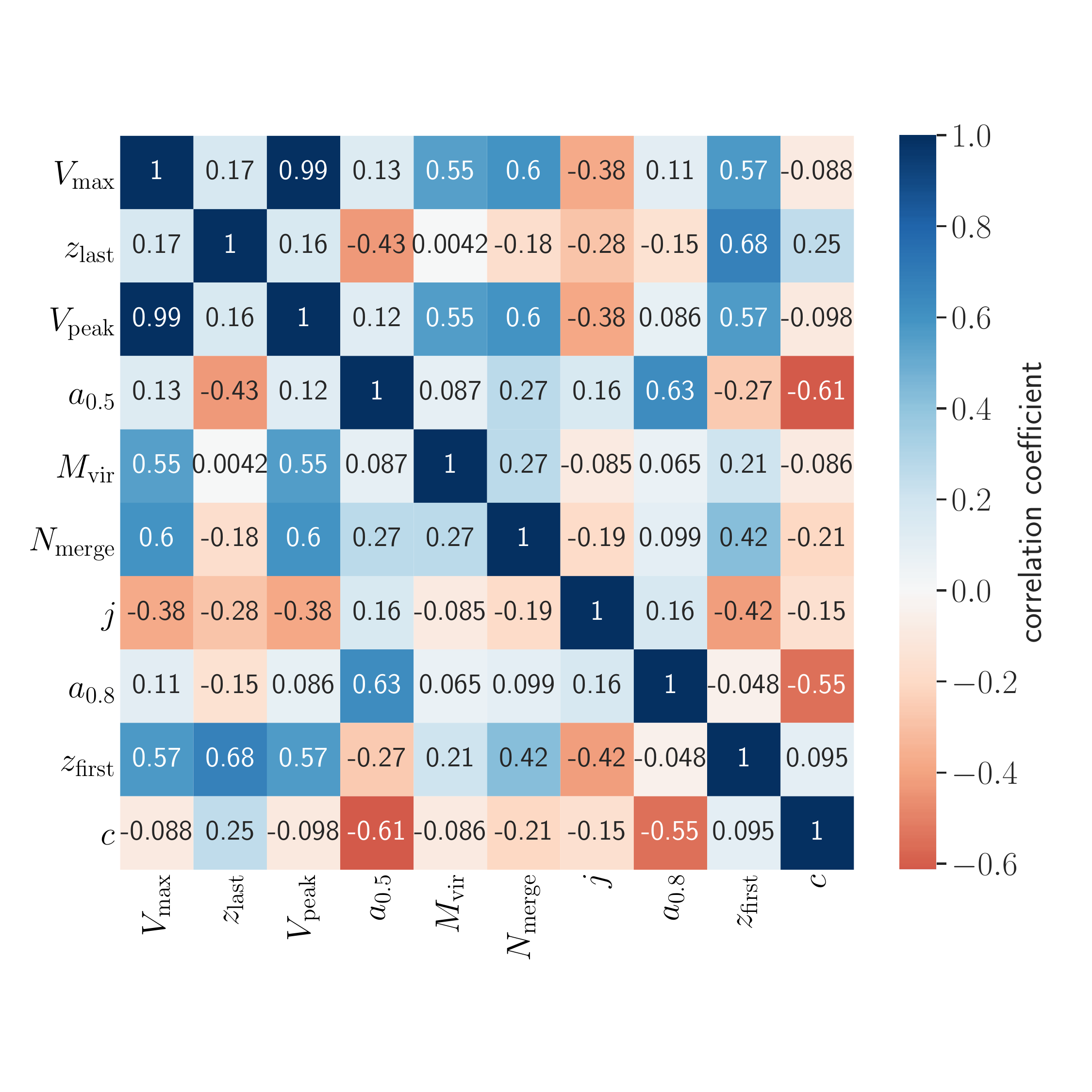}
	\end{subfigure}
	\hfill
\caption{
\textbf{Left:} Relative feature importance for the top 10 features of the RF predictions for central galaxies. \textbf{Right:} The correlation matrix of these top 10 features. The numbers shown are Pearson correlation coefficients between each pair of features. 
}
\label{feature_imp_cent}
\end{figure*}

\begin{figure*}
	\centering
	\begin{subfigure}[h]{0.46\textwidth}
		\includegraphics[width=\textwidth]{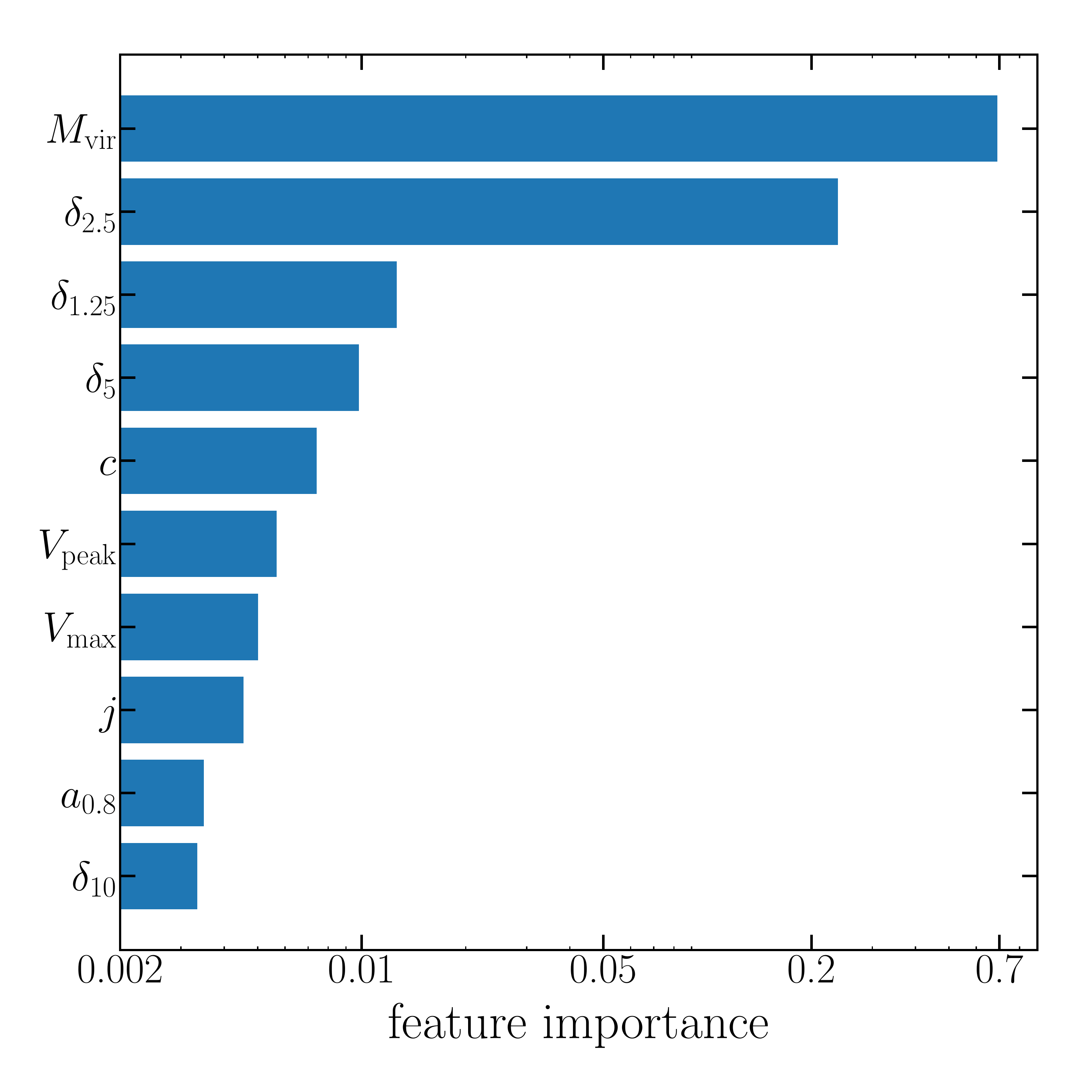}
	\end{subfigure}
	\hfill
	\begin{subfigure}[h]{0.52\textwidth}
		\includegraphics[width=\textwidth]{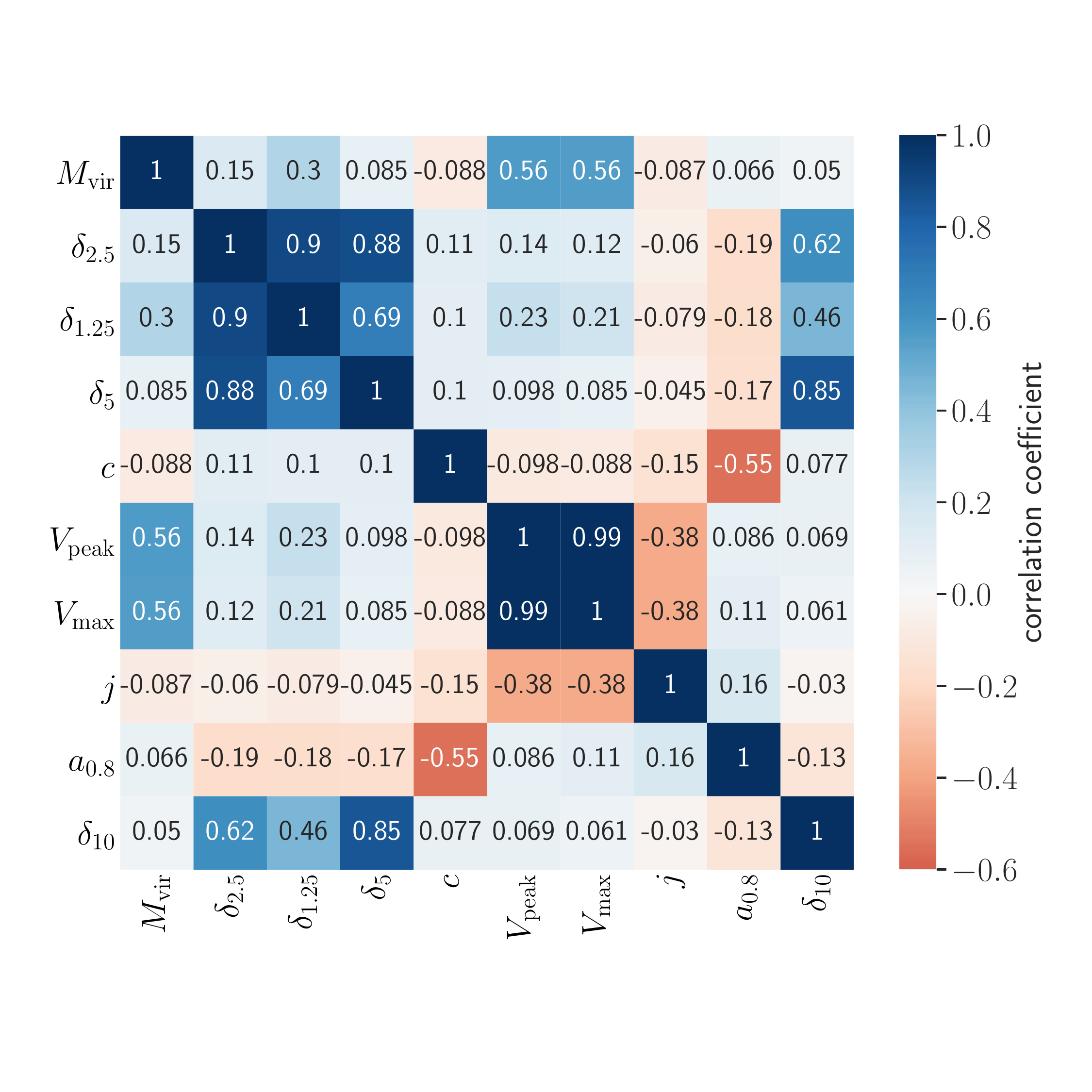}
	\end{subfigure}
	\hfill
\caption{
Relative feature importance and correlation matrix for the top 10 features of the RF predictions for satellite galaxies. 
}
\label{feature_imp_sat}
\end{figure*}

The above results show that the RF models are capable of accurately reproducing galaxy clustering and GAB. However, the number of input features is large which increases the complexity and running time of RF models. In this section, we aim to build simpler RF models with fewer input features that can achieve the same purpose. In addition to the prediction of galaxy numbers per halo, the RF algorithm also provides an estimate of the relative importance of the input features (i.e., all the secondary halo and environmental properties). It is evaluated based on the contribution of the input features to the construction of the RF decision trees. We show the top 10 properties ranked by feature importance in the left-hand side panels of Figure~\ref{feature_imp_cent} and Figure~\ref{feature_imp_sat}, for the central galaxies and satellites predictions, respectively.

For the central galaxies, we find that $V_{\rm max}$, the haloes' maximum circular velocity,  is the most important feature followed by $z_{\rm last}$, $V_{\rm peak}$, and $a_{\rm 0.5}$. $V_{\rm max}$ can be considered as a halo mass indicator (e.g., \citealt{Zehavi2019}), and the other properties characterise the formation history of a halo. $V_{\rm peak}$, the peak value of $V_{\rm max}$ over the history of the halo,  is a special case among them since it highly correlates with $V_{\rm max}$ (with a 0.99 Pearson correlation coefficient, as noted in the right panel of Figure~\ref{feature_imp_cent}). We perform a simple test that runs the RF prediction inputting the same feature twice (for example the halo mass), to mimic the situation of two highly correlated features). We find that it tends to maintain the importance of one feature and lower the importance of the other one. So it is likely that the roles of $V_{\rm max}$ and $V_{\rm peak}$ are comparable for the central galaxies prediction. Given the extreme correlation between the two, once $V_{\rm max}$ is utilized, $V_{\rm peak}$ does not really add any new information and thus it is not necessary to keep them both.

The importance of $V_{\rm max}$ is consistent with the finding by \citet{Zehavi2019} that $V_{\rm max}$ or $V_{\rm peak}$ better correlates with the central galaxies occupation than $M_{\rm vir}$ in the SAM sample, such that using the former reduces significantly the central galaxies OV with other secondary properties and the related trends in the stellar mass - halo mass relation. \citet{Xu2020} reach a similar conclusion with the Illustris simulation, namely that the stellar mass of central galaxies in fixed $V_{\rm peak}$ bins exhibits a weaker dependence on halo age or concentration than that in $M_{\rm vir}$ bins. This is not surprising since $V_{\rm max}$ or $V_{\rm peak}$ contains more internal structure information than $M_{\rm vir}$ alone, and in particular is also related to the concentration. Recently, \citet{Lovell2021} provide a ML approach to predict several galaxy properties from subhalo properties based on hydrodynamic simulations, also finding that $V_{\rm max}$ is the most important property for the prediction.

The next two properties in order of feature importance are $z_{\rm last}$ and $a_{\rm 0.5}$. Both are specific epochs in the formation history of the host halo. The halo formation time, $a_{\rm 0.5}$, is defined as the scale factor at the time when the host halo first reached half of its present mass, so is indicative of the halo age and is widely explored in assembly bias studies. At fixed halo mass, early-formed haloes (smaller $a_{\rm 0.5}$) tend to host more massive central galaxies than late-formed haloes (larger $a_{\rm 0.5}$), and thus are more likely to host central galaxies above a given stellar-mass threshold \citep{Zehavi2018}. The other parameter, $z_{\rm last}$, is the redshift of the last major merger of the host halo. It is another important epoch in the mass assembly history that could relate to the formation of the central galaxy. So it is reasonable that it is important for the central galaxies ML prediction. 
Interestingly, we find that no environmental properties appear in the top 10 features for central galaxies. This may be supported by the fact that the OV with environment is much smaller than with internal halo properties like age \citep{Zehavi2018}, as also demonstrated in Figure~\ref{ov_all}. However, recent studies have shown that environment is the most informative property for describing GAB \citep{Hadzhiyska2020,Xu2021}.  We will provide tests in the following subsections investigating the importance and necessity of the environment for reproducing the central galaxies and full GAB.

The left panel of Figure~\ref{feature_imp_sat} shows the feature importance for the satellites prediction. Halo mass, $M_{\rm vir}$, is the most important feature followed by the environment features $\delta_{\rm 2.5}$, $\delta_{\rm 1.25}$, and $\delta_{\rm 5}$. As expected, these three environmental measures are strongly correlated with each other, as can be seen in the right panel of Figure~\ref{feature_imp_sat}. In contrast to the central galaxies prediction, we note that here the environment is more important than secondary internal halo properties for predicting the number of satellites. Halo concentration is next and $V_{\rm max}$ and $V_{\rm peak}$ follow but with lower importance, which is again consistent with \citet{Zehavi2019} who showed that using $V_{\rm max}$ (or $V_{\rm peak}$) is detrimental to encapsulating the satellites OV relative to using $M_{\rm vir}$. These differences of feature importance between the central galaxies and satellites occupations highlight again the complexities of assembly bias. They imply that the formation and evolution of central and satellite galaxies may follow different paths and are impacted by different internal or environmental halo properties, and it is reasonable to model them separately with machine learning.

While the RF model estimates the input features importance, we should keep in mind that the features are correlated with each other. We take this into consideration when attempting to select fewer features for a less complex model. To illustrate that, in the right panel of Figure~\ref{feature_imp_cent} and Figure~\ref{feature_imp_sat}, we plot the correlation matrix which shows the Pearson correlation coefficients between each pair of the top 10 features included in the left panels. A correlation coefficient of 1 (shown by dark blue) indicates a positive maximal one-to-one correlation between the two properties, and a correlation coefficient of -1 (shown by dark orange) indicates a maximal anti-correlation. A correlation coefficient close to 0 indicates no correlation, with the two properties largely independent of each other. Values between 0 and 1 (-1) represent then a positive (negative) correlation with scatter, and the scatter is smaller for larger absolute values indicating a tighter correlation. In selecting a subset of top features, it is more effective to select a few such features that are important and yet less correlated with each other, in order to represent most of the information. For central galaxies, since $V_{\rm max}$ and $V_{\rm peak}$ are tightly correlated, we select $V_{\rm max}$, $z_{\rm last}$, $a_{\rm 0.5}$, and $M_{\rm vir}$ as the top features. For the satellite galaxies, we select $M_{\rm vir}$, $\delta_{\rm 2.5}$, $\delta_{\rm 1.25}$, and concentration $c$ as the top features. We show in the next section the RF prediction results with the selected top four features.

\subsection{Top Features}
\label{topfeatures}

\begin{figure*}
	\centering
	\begin{subfigure}[h]{0.8\textwidth}
		\includegraphics[width=\textwidth]{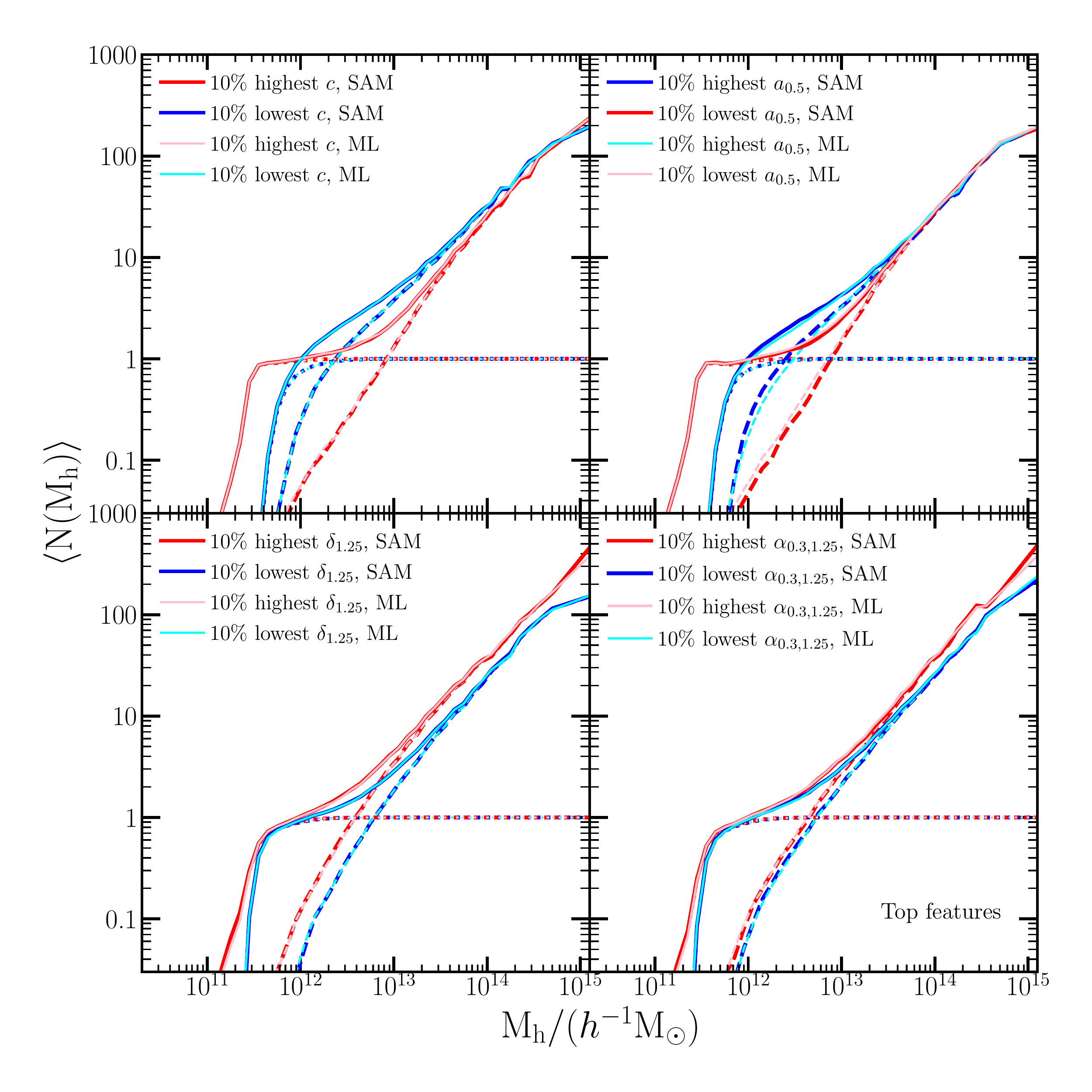}
	\end{subfigure}
	\hfill
\caption{Similar to Figure~\ref{ov_all}, the predicted OV with $c$, $a_{\rm 0.5}$, $\delta_{\rm 1.25}$, and $\alpha_{\rm 0.3,1.25}$, but now when using only the top four features in the RF algorithm. The four features for the central galaxies are $V_{\rm max}$, $a_{\rm lastmerg}$, $a_{\rm 0.5}$, and $M_{\rm vir}$. The four features for the satellite galaxies are $M_{\rm vir}$, $\delta_{\rm 2.5}$, $\delta_{\rm 1.25}$, and $c$. 
}
\label{ov_top}
\end{figure*} 

\begin{figure*}
	\centering
	\begin{subfigure}[h]{0.48\textwidth}
		\includegraphics[width=\textwidth]{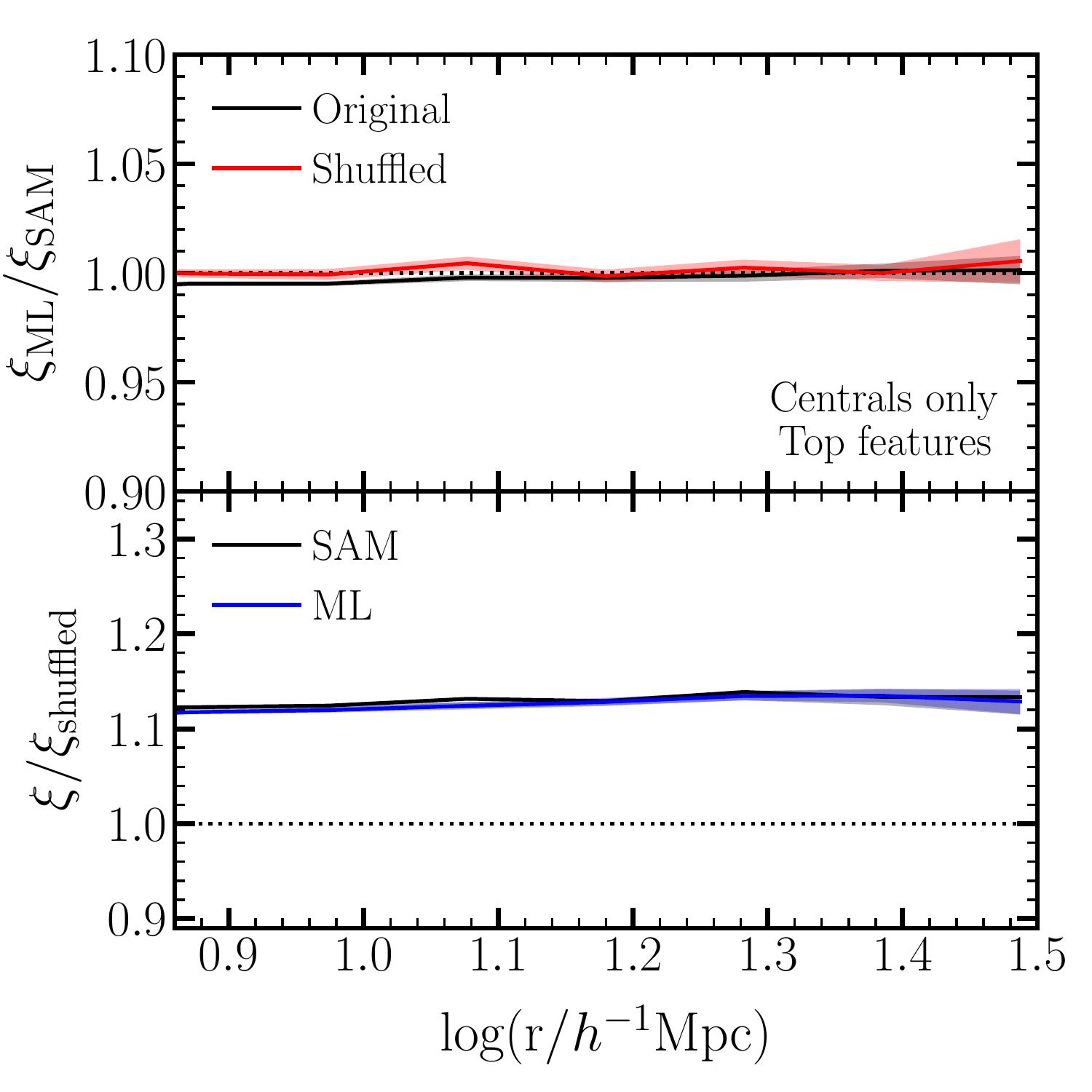}
	\end{subfigure}
	\hfill
	\begin{subfigure}[h]{0.48\textwidth}
		\includegraphics[width=\textwidth]{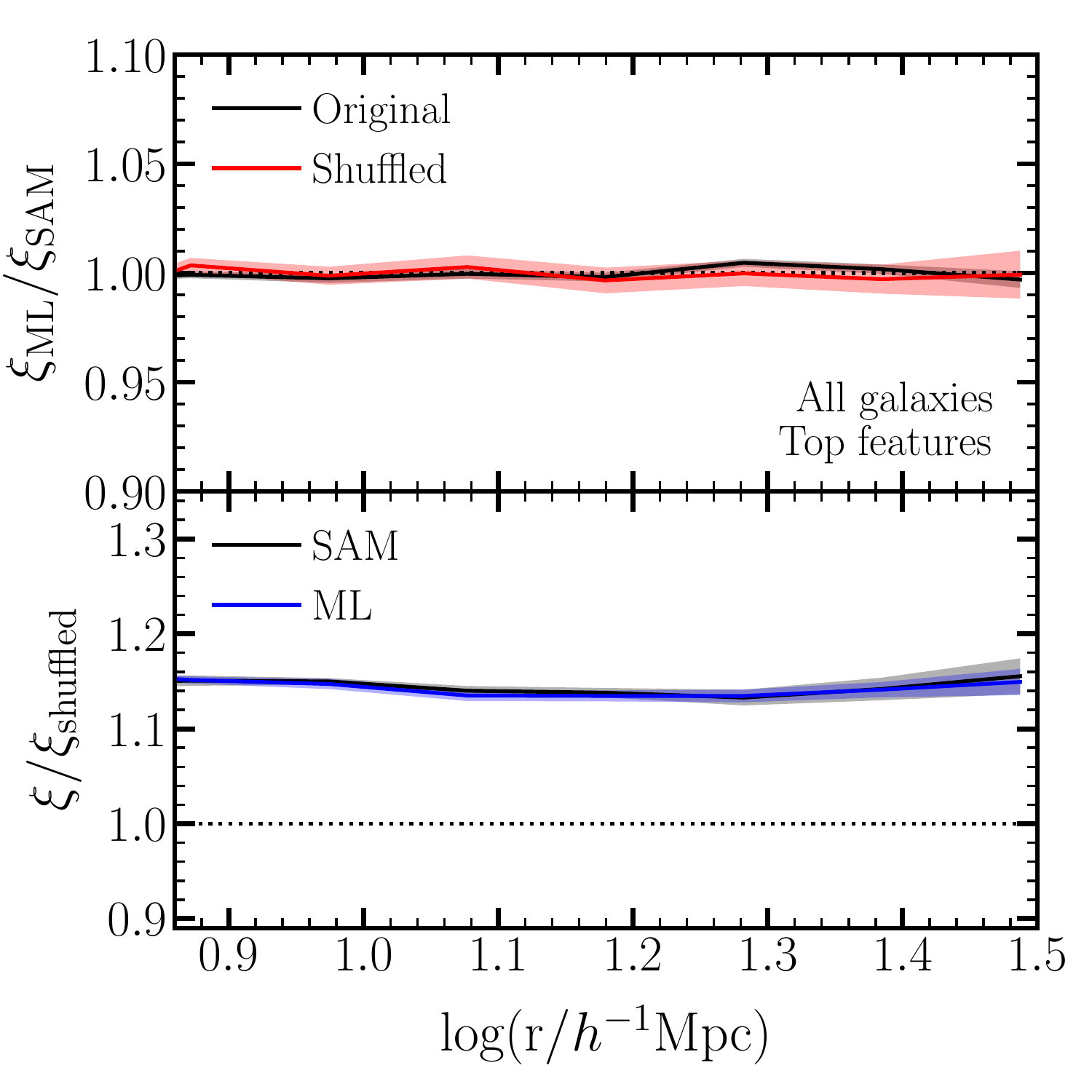}
	\end{subfigure}
	\hfill
\caption{
Similar to Figure~\ref{cluster_gab_all}, the predicted galaxy clustering and GAB measurement for centrals only (left) and all (central and satellite) galaxies (right), now obtained using only the top four features for central galaxies and satellites in the ML.
}
\label{cluster_gab_top}
\end{figure*}     

In this section, we predict the number of central and satellite with the top four features selected separately for central galaxies and satellites in Section~\ref{feature_imp}. We first perform new grid searches for the two sets of top features to tune the RF classification and regression models for centrals and satellites, respectively. The $F_1$ and $R^2$ scores are listed in the third and fourth lines of Table~\ref{table:clustering}, which are very similar to those from the all features models. Figure~\ref{ov_top} presents the ML predicted OVs compared to those from the SAM. Similar to the OV prediction with all features shown in Figure~\ref{ov_all}, the considered OVs are all accurately reproduced. This is even more impressive in this case, when using only four features for each centrals or satellites. It is worth noting that other than $M_{\rm vir}$ which is common to both, no secondary property is present in both the centrals and satellites top features. Thus in all panels of Figure~\ref{ov_top}, showing the OV with $c$, $a_{\rm 0.5}$, $\delta_{\rm 1.25}$, and $\alpha_{\rm 0.3,1.25}$, these properties are not involved in all predictions. We therefore conclude that the top four features for centrals and satellites are highly efficient in capturing the information needed for reproducing the halo occupation numbers.

With the predicted occupations from the top features, we again populate the haloes to create a mock galaxy catalogue and measure galaxy clustering and the GAB signal. The results are presented in Figure~\ref{cluster_gab_top} and summarized in Table~\ref{table:clustering}. For the centrals-only sample, the predicted original clustering, shuffled clustering, and the GAB are highly consistent with those of the SAM (left panels), with recovered fractions of 1.00, 1.00, and 0.97, respectively. These results are very similar to those from the prediction using all features, and the RF classification with the top four features works equally well as the one with all features. It is worth noting again that the top four features for central galaxies are all halo internal properties without explicitly including environment. This seems to imply that environment measures are not necessary for recovering the centrals GAB. However, other works have shown that environment is crucial for capturing GAB (\citealt{Hadzhiyska2020,Xu2021}; C.\ Cuesta, in prep.). To gain more insight on the role of environment in recovering GAB, we examine in Section~\ref{mass_env} obtaining ML predictions based on only mass and environment, and in Section~\ref{internal} the predictions based solely on internal halo properties.

The right-hand side panels of Figure~\ref{cluster_gab_top} provide the predicted clustering and GAB for all galaxies including satellites. The satellite occupation is predicted with the top four features specific for satellites selected in Section~\ref{feature_imp} (which are different than the top four features for centrals) and include environmental properties. The recovered original clustering, shuffled clustering, and GAB fraction are all in excellent agreement with the SAM measurements (with recovered fractions of 1.00 for all). These fractions are in fact slightly higher than those for the all features model, but we consider them to be equally good due to the randomness associated with the prediction, populating galaxies, and shuffling. Combining the results from the centrals and satellites predictions, we find that the ML mock with only the top four features for each can well capture the galaxy-halo connection in the SAM and reproduce the expected galaxy clustering and GAB.

\subsection{Halo mass and one environmental feature}
\label{mass_env}

\begin{figure*}
	\centering
	\begin{subfigure}[h]{0.48\textwidth}
		\includegraphics[width=\textwidth]{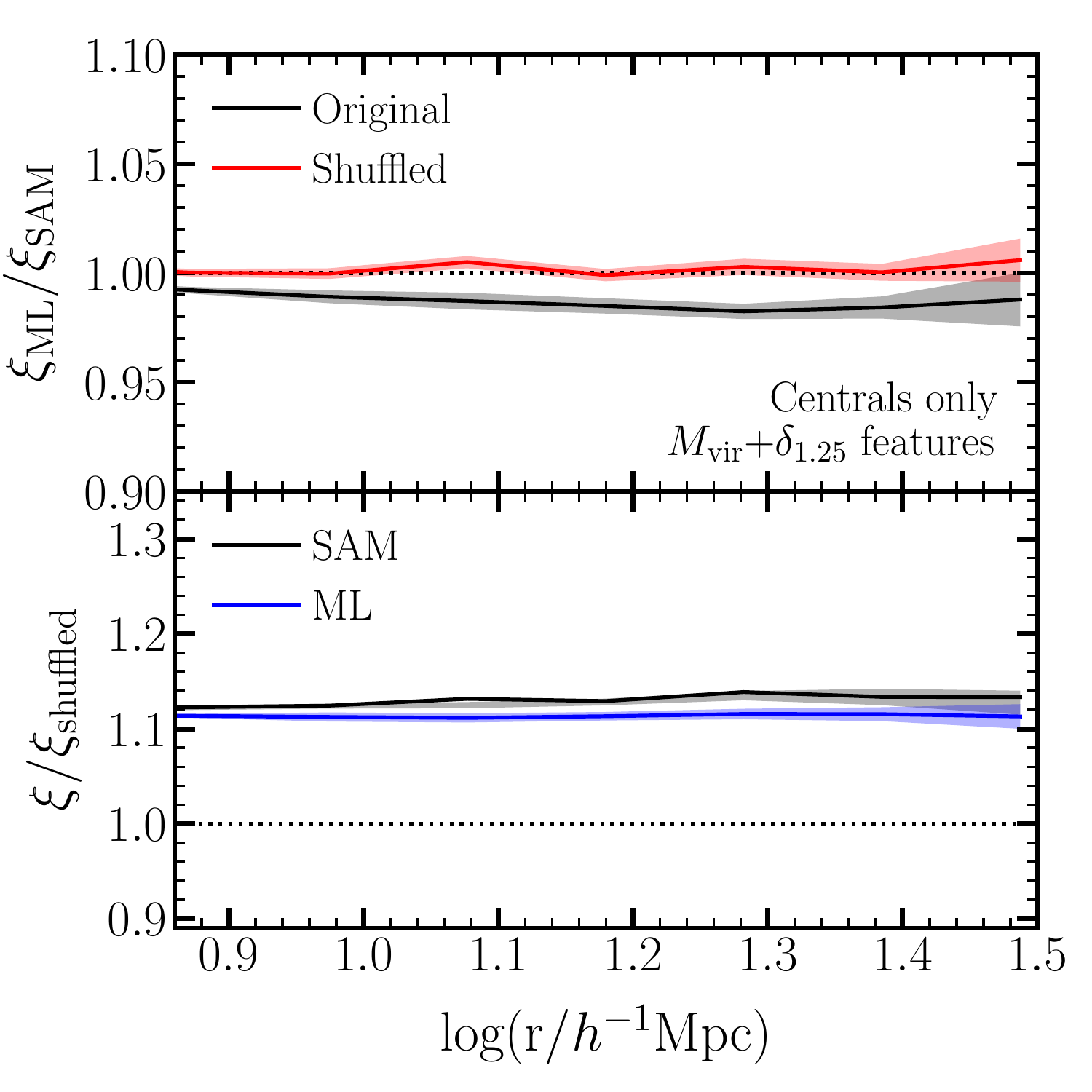}
	\end{subfigure}
	\hfill
	\begin{subfigure}[h]{0.48\textwidth}
		\includegraphics[width=\textwidth]{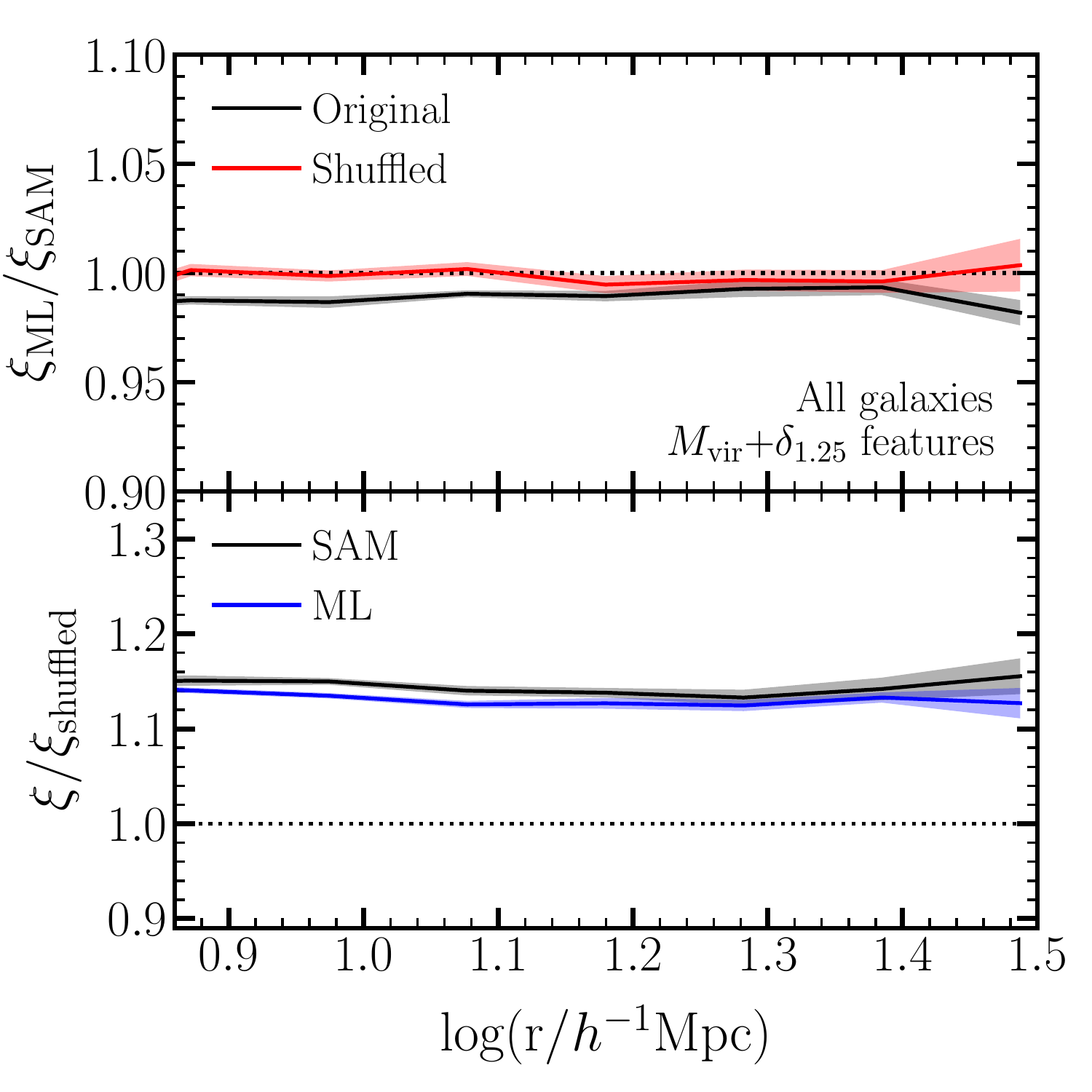}
	\end{subfigure}
	\hfill
\caption{
Similar to Figure~\ref{cluster_gab_top}, the predicted galaxy clustering and GAB measurement for centrals only (left) and all galaxies (right), using only halo mass and environment $\delta_{\rm 1.25}$ for both centrals and satellites. 
}
\label{cluster_gab_mass_env}
\end{figure*}  

In Section~\ref{feature_imp} and Section~\ref{topfeatures}, we saw that environmental properties are listed in the top features for the satellite galaxies occupation. However, they are not included in the top 10 features for the central galaxies prediction, and the top four features for centrals (without environment) can well reproduce the centrals GAB. This seems to suggest that environment is not necessary for a recovery of the centrals GAB. We clarify that the internal halo properties (e.g., age $a_{\rm 0.5}$) are surely dependent on environment to some degree, since they produce assembly bias,  but it is of interest to know whether an environmental measure is needed to be explicitly included. Traditional (non-ML) analyses show that environment is the most informative property for GAB, more significant than any other single secondary property in either SAM or hydrodynamic galaxy samples \citep{Hadzhiyska2020,Hadzhiyska2021,Xu2021}. In particular, \citet{Xu2021} demonstrated that $\delta_{\rm 1.25}$ can capture the full level of GAB in the SAM. To further examine the role of environment in GAB, we repeat our analysis but now only use the halo mass $M_{\rm vir}$ and $\delta_{\rm 1.25}$ as input features to the RF algorithm models.

The OVs predicted by the ML models based on $M_{\rm vir}$ and $\delta_{\rm 1.25}$ are shown in Appendix~\ref{ov_ME}. We find that the models are less successful in reproducing the OVs compared to the models with all features and the top four features. The OV dependence on $\delta_{\rm 1.25}$ is recovered as expected, as well as the ones for $\alpha_{\rm 0.3,1.25}$ to a large extent. However, the variations with halo properties such as concentration and age are poorly recovered, especially for the satellites.  We note that these results are in agreement with those by \citet{Xu2021}. While they were able to mimic the full level of GAB with only halo mass and $\delta_{\rm 1.25}$, they were similarly unable to recover the OVs with other secondary properties. In our ML analysis, the weaker recovery is also reflected by the somewhat lower $F_1$ and $R^2$ performance scores of the RF models in this case (lines 5-6 in Table~\ref{table:clustering}). These scores reflect the halo-by-halo prediction accuracy, such that a lower value will lead to less accurate recovery of the OVs.  

We then populate haloes with the predicted occupations and measure galaxy clustering and GAB. The results for these are shown in Figure~\ref{cluster_gab_mass_env} and summarized in Table~\ref{table:clustering} as well. For both centrals-only and all galaxies, the predicted shuffled clustering is in perfect agreement with the SAM results (the red solid lines in the top panels), indicating that the halo mass dependence of clustering is reproduced. However, for the original (unshuffled, including assembly bias) SAM clustering the ML recovery for both these cases is slightly lower (the black solid lines in the top panels). It is still reasonably good with a recovery fraction of 0.99, but stands out in contrast to the excellent agreement of the predictions with all features and top features explored earlier. This leads to a reduced ability to recover the GAB signal, denoted by the solid blue lines in the bottom panels of Figure~\ref{cluster_gab_mass_env}. These correspond to $f_{\rm AB}$ values of 0.86 and 0.92 for the centrals-only GAB and the all-galaxies one, respectively. This result is consistent with \citet{Xu2021} who show that shuffling galaxies among haloes with fixed mass and $\delta_{\rm 1.25}$ (which can also be considered as populating haloes according to only mass and $\delta_{\rm 1.25}$) reproduces $\sim$90\% of the full GAB signal. The performance of the RF models based on only mass and environment is also similar to that of the modified HOD model provided by \citet{Xu2021}, while in the latter the GAB parameters are tunable to reproduce the full effect.

Our analysis suggests that mass and environment are efficient in capturing most of the GAB signal and are useful for reproducing galaxy clustering within 1\% if halo internal properties are unavailable. Combined with the results from Section~\ref{topfeatures}, we find that the central GAB can be recovered with either a few internal halo properties or the environment. The former achieves the purpose by capturing most of the assembly bias effects in halo occupation, whereas the latter achieves this by ``mimicking'' the effect on the clustering. The satellites assembly bias effects can be largely recovered by environment alone, but including information on internal properties improves the OV. Would internal properties alone be able to reproduce both the centrals and satellites GAB? Is the environment required for reproducing the full GAB? We answer these questions in Section~\ref{internal} by testing the RF models using now only the internal halo properties.

\subsection{Internal Features}
\label{internal}

\begin{figure*}
	\centering
	\begin{subfigure}[h]{0.48\textwidth}
		\includegraphics[width=\textwidth]{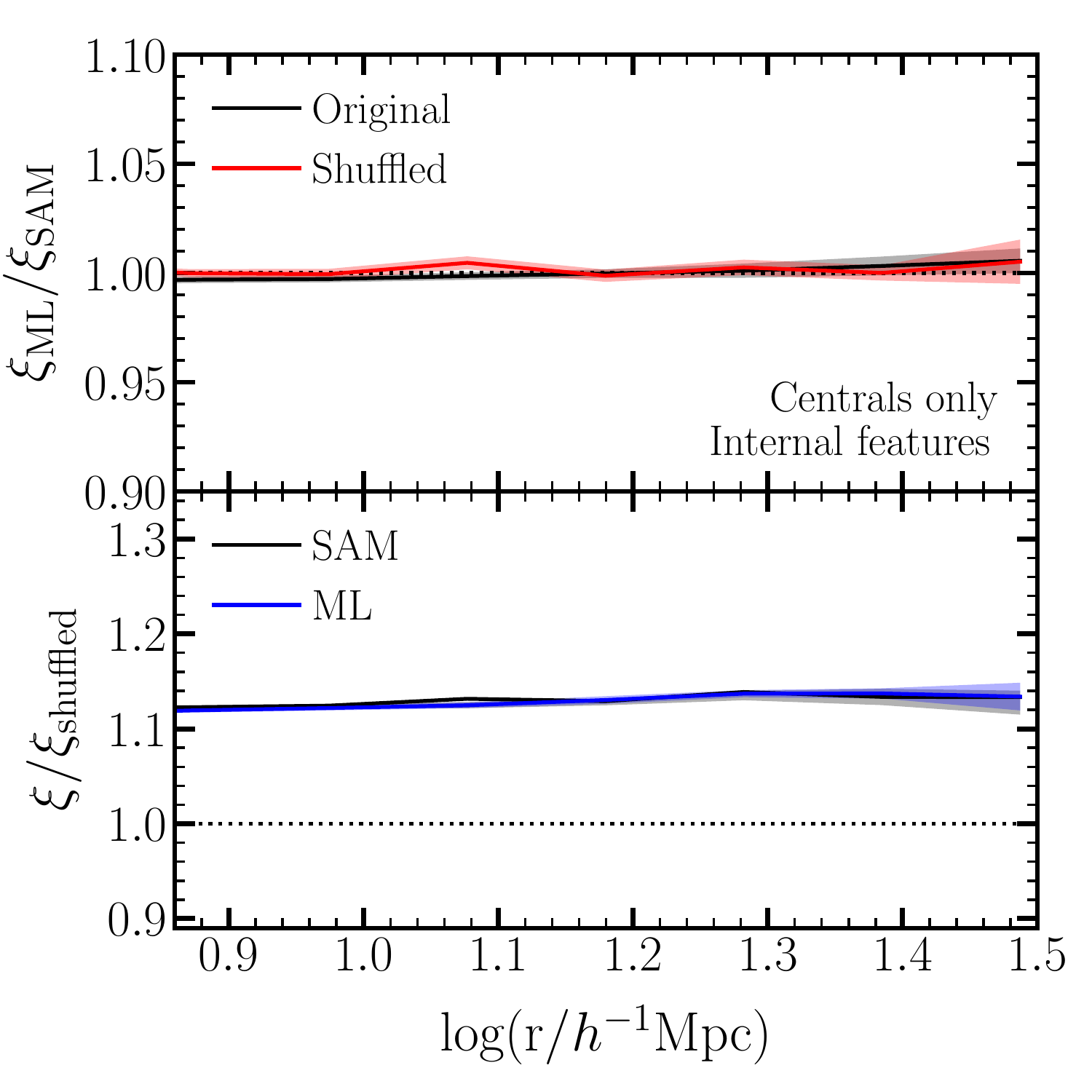}
	\end{subfigure}
	\hfill
	\begin{subfigure}[h]{0.48\textwidth}
		 \includegraphics[width=\textwidth]{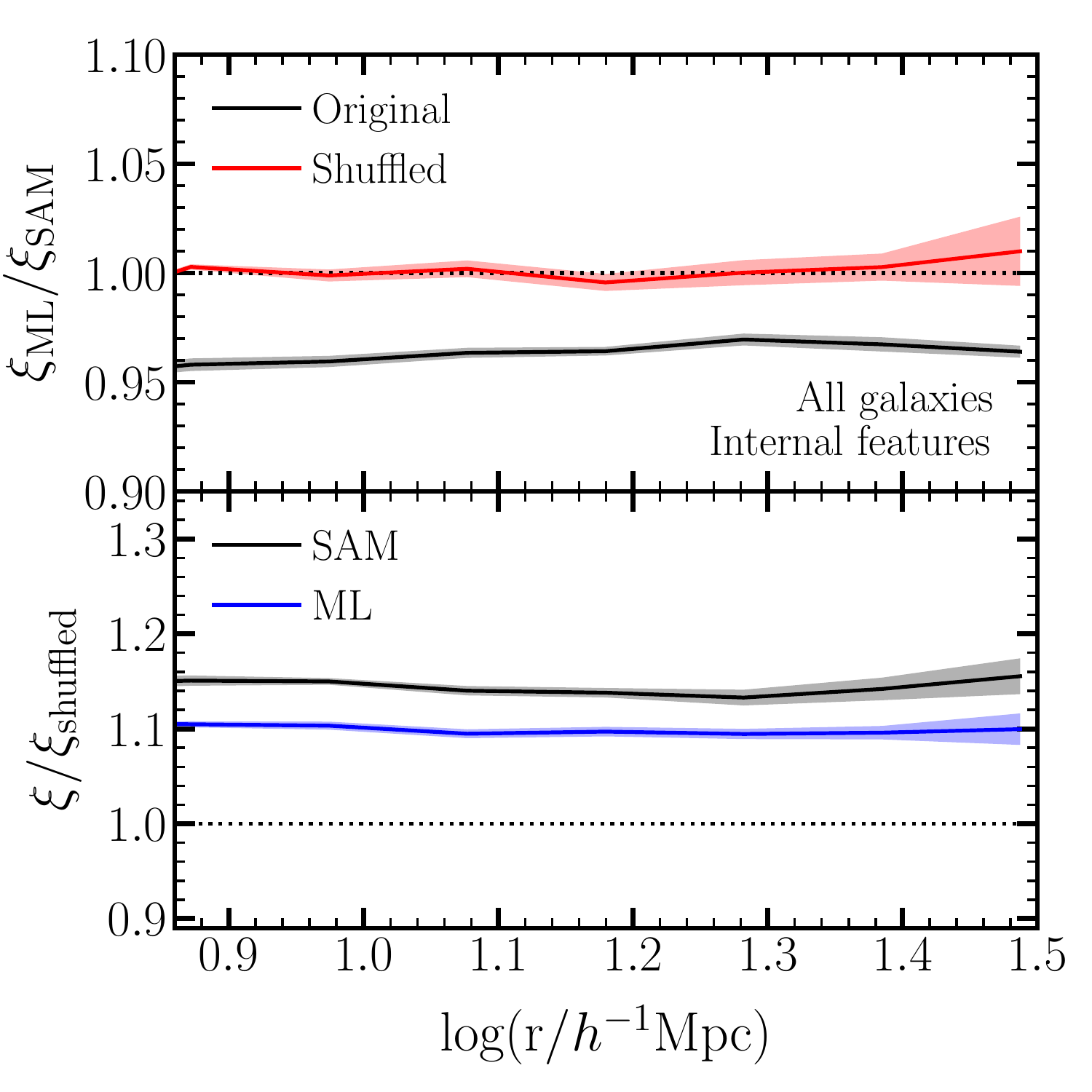}
	\end{subfigure}
	\hfill
\caption{
Similar to Figure~\ref{cluster_gab_all}, the predicted galaxy clustering and GAB for centrals only (left) and all galaxies (right) when using all internal properties (and no environment measures) for the ML predictions.  
}
\label{cluster_gab_internal}
\end{figure*}  

In this section, we explore the performance of the ML predictions when using only internal halo properties, commonly associated with halo assembly bias, rather than environment measures directly.  We include all the halo properties listed in lines 1-13 of Table~\ref{table:haloprops}. In contrast to the previous case with halo mass and environment, the models with internal properties accurately recover the OV with concentration and $a_{\rm 0.5}$ accurately, as shown in Figure~\ref{ov_in} in Appendix~\ref{more_ov}. The OV with $\delta_{\rm 1.25}$ and $\alpha_{\rm 0.3,1.5}$ are partially recovered, with the centrals OV well reproduced but smaller OV for the satellite galaxies. As before, we proceed to create mock galaxy catalogues with the ML predicted occupations, to study the impact on clustering and GAB.

The clustering and GAB of the RF mock are shown in Figure~\ref{cluster_gab_internal}. For the central galaxies only (left-hand side), we find that the original clustering, shuffled clustering, and GAB are all well reproduced at sub percent accuracy. These results are similar to those with only the top four properties shown in Section~\ref{topfeatures}, which for the central galaxies were comprised of only internal properties ($V_{\rm max}$, $z_{\rm last}$, $a_{\rm 0.5}$, and $M_{\rm vir}$). These top properties appear to include most of the information needed to reproduce the centrals clustering and GAB, such that now including all internal properties does not change the results. 
The situation for the satellites, however, is different since environment measures have a prominent role in the top features.  Consequently, we find that when adding the satellite galaxies, the clustering and GAB are not well with only the internal properties. The recovered clustering is lower than that of the SAM by 3\%, and only 70\% of the GAB is reproduced.

We conclude that while the environment is not necessary for centrals clustering, it is required for an accurate representation of the satellites clustering. This is consistent with the feature importance provided by the RF models. In summary, secondary halo properties include enough information to recover in full the centrals OV, clustering and GAB, but the environment is needed for accurately predicting the satellites OV and the full level of clustering and GAB, and cannot be replaced with internal properties alone.

\subsection{Single-Epoch Features}
\label{nontree}

\begin{figure*}
	\centering
	\begin{subfigure}[h]{0.48\textwidth}
		\includegraphics[width=\textwidth]{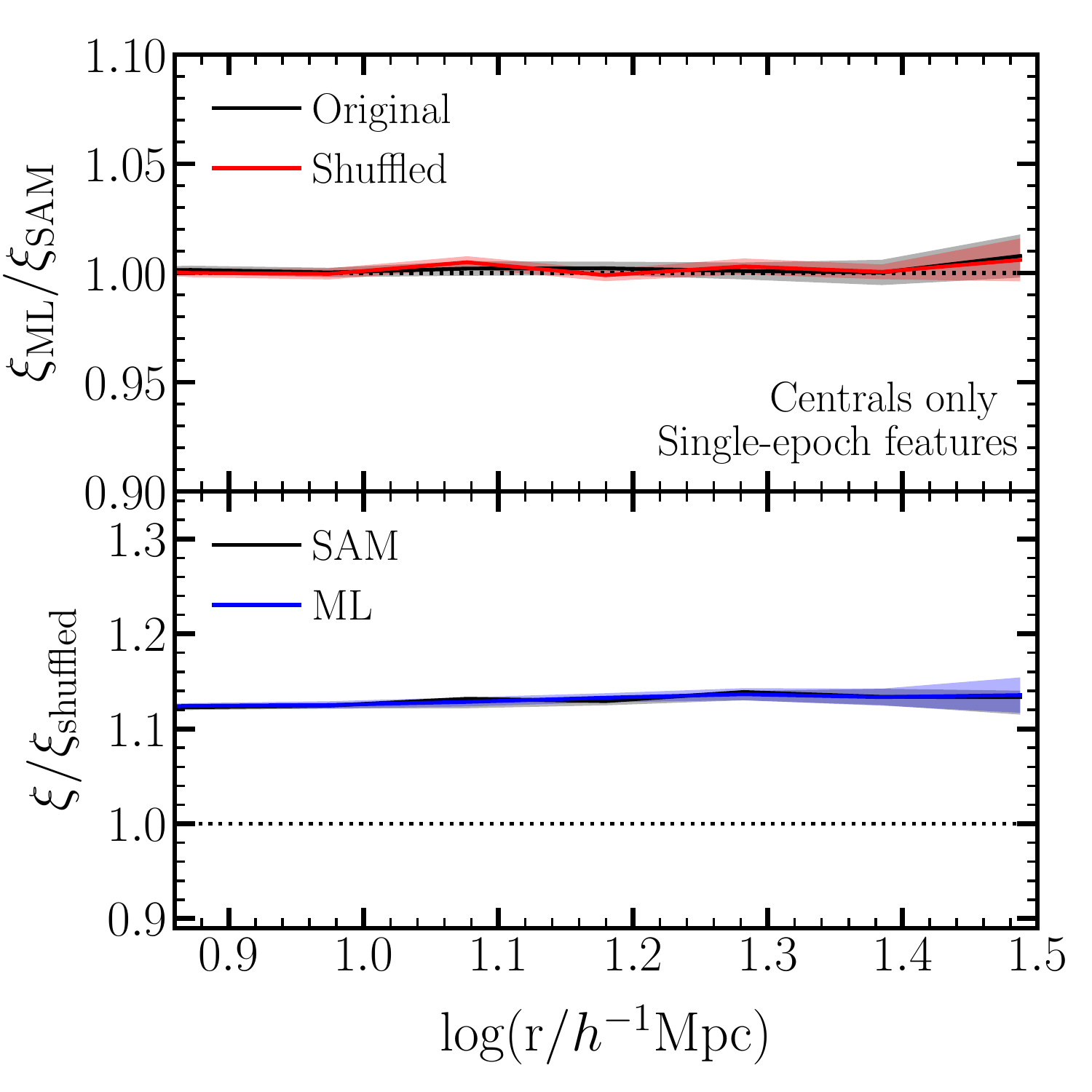}
	\end{subfigure}
	\hfill
	\begin{subfigure}[h]{0.48\textwidth}
		 \includegraphics[width=\textwidth]{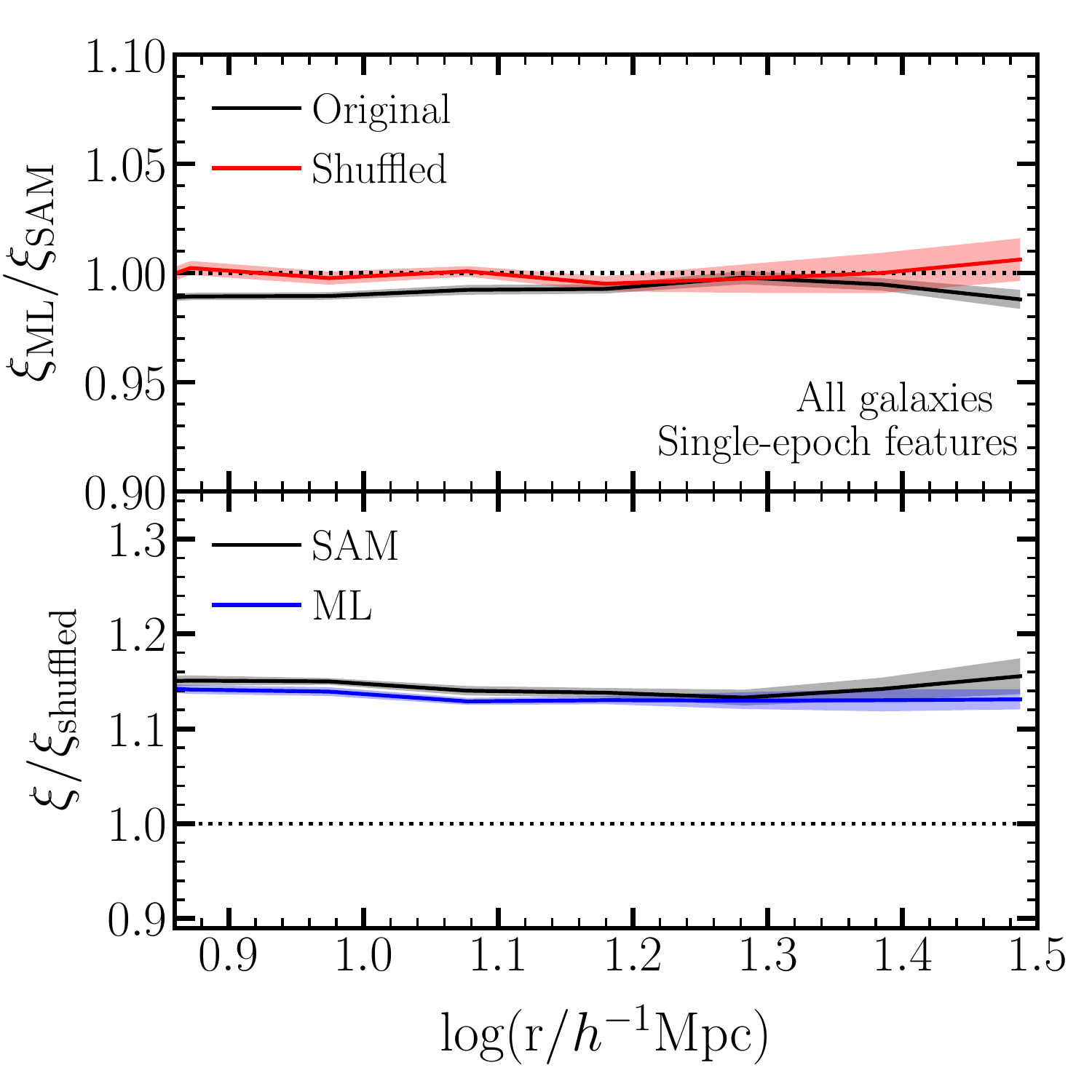}
	\end{subfigure}
	\hfill
\caption{
Similar to Figure~\ref{cluster_gab_internal}, the predicted galaxy clustering and GAB for centrals only (left) and all galaxies (right), now using only the following single-epoch (i.e not involving the halo merger tree) features $M_{\rm vir}$, $V_{\rm max}$, concentration $c$, angular momentum $j$, and $\delta_{\rm 1.25}$for both centrals and satellites. 
}
\label{cluster_gab_nontree}
\end{figure*}

The main purpose of this paper is to explore the possibility of creating realistic mock galaxy catalogues from halo catalogues of $N$-body simulations using ML to capture the detailed galaxy-halo connection. However, for some low-resolution $N$-body simulations, the halo merger tree which follows the haloes' evolution is unavailable. In such cases, one will not be able to obtain halo properties that rely on the merger tree, such as $a_{\rm 0.5}$, $V_{\rm peak}$, and $z_{\rm last}$. The only available properties will be single-epoch properties typically obtained from the final snapshot of the simulation.  These include $M_{\rm vir}$, $V_{\rm max}$, concentration $c$, angular momentum $j$, and the environment measures. In this section we test the performance of ML models based on these single-epoch properties. We include, for both centrals and satellites, the above four internal halo properties and $\delta_{\rm 1.25}$.

We find that the OVs in this case are mostly well reproduced, as shown in Figure~\ref{ov_non}. The OV with $c$ and $\delta_{\rm 1.25}$ are particularly well reproduced, as expected, since they are part of the input features.  The only notable deviation is for the centrals OV with $a_{\rm 0.5}$ where the ML prediction is slightly smaller than in the SAM. The predicted clustering and GAB signal are shown in Figure~\ref{cluster_gab_nontree}. For the centrals-only prediction, both galaxy clustering and GAB are extremely well reproduced. Adding the satellites, the SAM clustering is recovered to within 1\% and the GAB is recovered to within 5\%. These are better than the ML with only internal properties or $M_{\rm vir}$ and $\delta_{\rm 1.25}$ alone, and slightly worse than the models with all features or the top four features. We suspect that including additional available (single-epoch) environment measures, such as $\delta_{\rm 2.5}$, would have improved this result.

Overall, the analysis illustrates that when the halo formation history is not available (for example, in low-resolution $N$-body simulations), ML models can still reproduce the clustering and GAB to reasonable accuracy. Using ML to predict the halo occupation and populate haloes with galaxies accordingly thus provide a viable practical approach to creating realistic mock galaxy catalogues, even in such cases.

\section{Summary and Discussion}
\label{summary}

In this paper, we describe a machine learning approach to predict the number of galaxies above a stellar-mass threshold in dark matter haloes using halo and environment properties as input. We use the halo catalogue from the Millennium simulation and the galaxy sample from the \citet{Guo2011} SAM model to train and test our ML method. We use random forest classification and regression for the central galaxies and satellites, respectively, and adopt commonly-used $F_1$ and $R^2$ scores to evaluate the performance of the models. We test different combinations of input properties. For each set of the input properties, we tune the hyper-parameters of the RF models to maximize the performance scores. With the predicted number of central and satellite galaxies in each halo, we then populate the Millennium simulation haloes to create a mock galaxy catalogue and measure the galaxy clustering and galaxy assembly bias signal to compare with those of the SAM.

We start by using all the available internal and environmental halo properties, listed in Table~\ref{table:haloprops}, as input features. The predicted HOD and occupancy variations are consistent with those measured from the SAM. The ML mock catalogue matches well the galaxy clustering, shuffled sample clustering, and GAB as that of the original SAM sample. The clustering is recovered to sub percent accuracy and GAB is recovered at the two percent level. Our results show that machine learning is capable of capturing the complex high-dimensional relations between halo properties and the galaxy occupation in the SAM model and reproduce the expected galaxy clustering accurately, including the intricate effects of assembly bias.

The RF models also provide an estimate of the relative importance of the different features. We find that $V_{\rm max}$ is the most important feature for central galaxies, followed by formation history (internal) properties and halo mass. Environmental properties are not included in the top 10 features. On the other hand, the satellite galaxies prediction relies the most on halo mass and environmental properties.  We construct simpler RF models with the top four halo properties for the centrals and satellite galaxies separately, based on the feature importance and correlation matrix between them.  We select $V_{\rm max}$, $z_{\rm last}$, $a_{\rm 0.5}$, and $M_{\rm vir}$ for central galaxies prediction and $M_{\rm vir}$, $\delta_{\rm 1.25}$, $\delta_{\rm 2.5}$, and concentration $c$ for the satellites prediction. The OVs, clustering and GAB are again well reproduced. This demonstrates that the ML methodology is powerful enough such that, with only a few halo properties, it can achieve similar performance as when using all the available information.

We perform two additional tests to further explore the role of the environment in reproducing galaxy clustering and GAB. We first use only halo mass and one environmental property ($\delta_{\rm 1.25}$) as input for the RF models. With the ML-constructed mock, we still recover the SAM (original and shuffled) galaxy clustering to within 1\% and about 92\% of the full GAB signal (Figure~\ref{cluster_gab_mass_env}). We conclude that $\delta_{1.25}$ along with the host halo mass is enough to reproduce GAB to $\sim 10\%$ accuracy. This is in agreement with previous works \citep{Hadzhiyska2020,Contreras2021,Xu2021} that showed that using environmental properties can realistically incorporate assembly bias into empirical models such as the HOD or SHAM. However, these methods do not recover the full occupancy variation for halo properties with inherent halo assembly bias, such as concentration or age (see Appendix~\ref{more_ov} for more details). This puts a limitation on such approaches when using statistics that need a more detailed modelling of the galaxy-halo connection (like galaxy lensing). Other approaches that add assembly bias to mock catalogues using a single secondary property like the halo concentration will also necessarily fail to reproduce the galaxy-halo connection, since such properties are not able to capture on their own the full GAB of a semi-analytic galaxy sample \citep{Croton2007,Xu2021}. To our knowledge, the approach presented in this paper is the most efficient model capable of populating galaxies in N-Body simulations, while taking into account the correlations between the halo occupation and the secondary halo properties, and recovering a realistic GAB signal.

The second test employs all secondary assembly bias properties as input, excluding the environment. The clustering and GAB for central galaxies alone are recovered at sub percent accuracy, at the same level as those with all or top four properties. However, after adding satellite galaxies, the predicted ML mock catalogue only recovers about 70\% of the GAB signal. This clearly indicates that internal properties alone are not able to fully capture the relation between the satellite occupations and the host haloes.  Perhaps further information can be introduced by including additional internal properties not included in this work, however using readily-available environment measures seems the more practical approach here. Combining the results from the two tests, we find that both internal properties and environmental properties can reproduce the centrals clustering and GAB, but that environment is necessary for reproducing the full clustering and GAB. Furthermore, environment alone (together with halo mass) goes a long way toward mimicking the correct level of assembly bias, however including assembly bias properties in needed to recover the OV with such properties and reproduce GAB to percent level accuracy.

Finally, to explore a potential application of our ML method in cases where the halo merger tree might not be available in low-resolution $N$-body simulations, we limit the input properties to single-epoch ones which can be obtained from the present-day simulation. We therefore use $M_{\rm vir}$, $V_{\rm max}$, concentration, angular momentum, and $\delta_{\rm 1.25}$ as input for the RF models. The OVs in this case are reasonably reproduced, galaxy clustering is matched at sub percent level, and the GAB signal is recovered to 5\%. An improvement in the GAB level may be reached if including additional environment parameters. Utilizing such a model can be a practical approach for populating large dark-matter-only simulations, like the Millennium XXL Simulation \citep{Angulo2012} and others, where the resolution of the halo merger trees is insufficient for use in a SAM. Instead, one can train and fine-tune a ML model on a smaller volume high-resolution galaxy formation simulation. Once the model is determined, it is straight-forward to apply it to the larger simulation to create mock galaxy catalogues with all the required attributes.

Overall, our results demonstrate the ability of machine learning to successfully capture the high-dimensional relationship between the halo occupation and multiple halo properties. Our tests here are with a SAM, but we expect similar performance when matching hydrodynamical simulations, which we leave for future work. As just mentioned, it is particularly advantageous to learn these relations from existing SAM or hydrodynamic galaxy samples in order to create realistic mock galaxy catalogues with haloes in larger cosmological volumes. This has the advantage of reproducing the detailed galaxy-halo connection of state-of-the-art galaxy formation models, which might be computationally-prohibitive otherwise. Additionally, with the single-epoch test, we show that ML can also be used to reproduce galaxy clustering and assembly bias in low-resolution $N$-body simulations for emulators, which are becoming benchmarks for cosmological studies. In this work, we focus on predicting the occupation of galaxies in halo for stellar-mass selected samples, but it can be extended to other types of galaxy samples, for example, star formation rate selected samples and colour selected sample which are also frequently used in observations, as well as galaxy samples at higher redshifts. We leave these as well for future studies.

Different studies in the literature have focused on predicting galaxy properties from haloes with ML techniques. \citet{Xu2013} predict the number of galaxies based on six halo properties and reproduce the galaxy clustering to a 5\%-10\%, which is similar to our internal properties predictions without using environment. Our extended work now reaches sub percent accuracy. Other works based on ML techniques predict properties of central galaxies such as stellar mass, star formation rate, and gas mass to mimic galaxy formation in hydrodynamic simulations (e.g., \citealt{Kamdar2016b,Agarwal2018,Wadekar2020}). In contrast, our study using the occupation number more directly probes galaxy clustering and assembly bias and allows to naturally predict both central and satellite galaxies. 

For the purpose of modelling the halo occupation, our work can be considered as a ML alternative to the HOD approach. The standard HOD framework models the number of galaxies in a halo as a function of only halo mass. Different extensions of the HOD (e.g., \citealt{Hearin2016,Xu2021,Yuan2021}) include an additional dependence on one or two secondary halo properties, but the galaxy-halo relations obtained are still limited. With ML-based methods, the non-linear dependence of the halo occupation on multiple halo and environment properties can be maximally reproduced, without assuming an analytic relation between them or fixing the parameters. Similarly, compared to empirical SHAM models, ML methods can capture and reproduce more complex multivariate dependencies between the galaxy and halo properties. This advantage makes ML a powerful approach for studying the galaxy-halo connection and for creating realistic mock galaxy catalogues which will be useful for upcoming large galaxy surveys.

\section*{Acknowledgements}
XX, SK and IZ acknowledge support by NSF grant AST-1612085. SC acknowledges the support of the ``Juan de la Cierva Formaci\'on'' fellowship (FJCI-2017-33816).

\section*{Data Availability}
The data underlying this article are available in GitHub at \url{https://github.com/xiaojux2020/RFmodels}



\appendix

\section{Results for other number densities}
\label{num_den}

\begin{figure*}
	\centering
	\begin{subfigure}[h]{0.48\textwidth}
		\includegraphics[width=\textwidth]{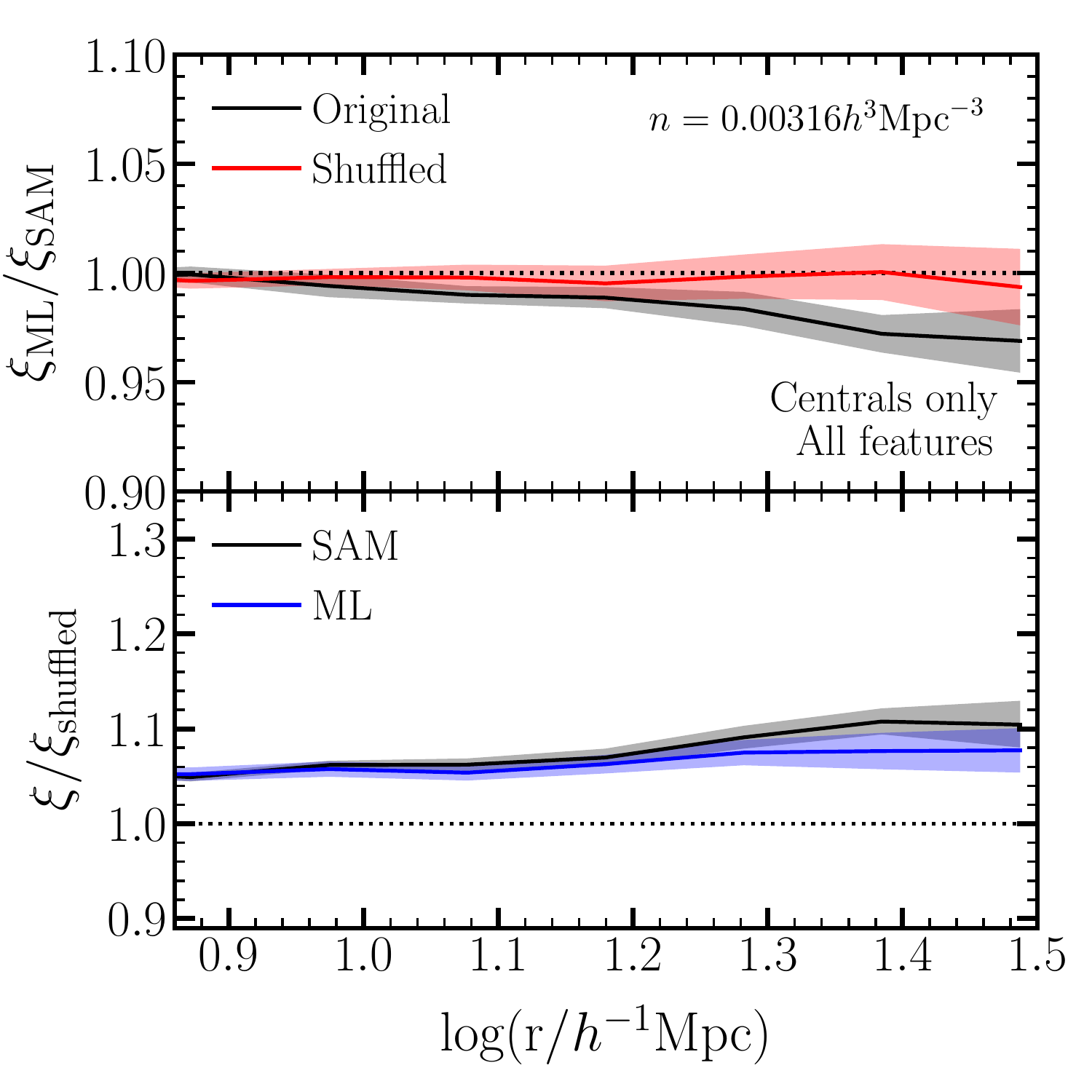}
	\end{subfigure}
	\hfill
	\begin{subfigure}[h]{0.48\textwidth}
        \includegraphics[width=\textwidth]{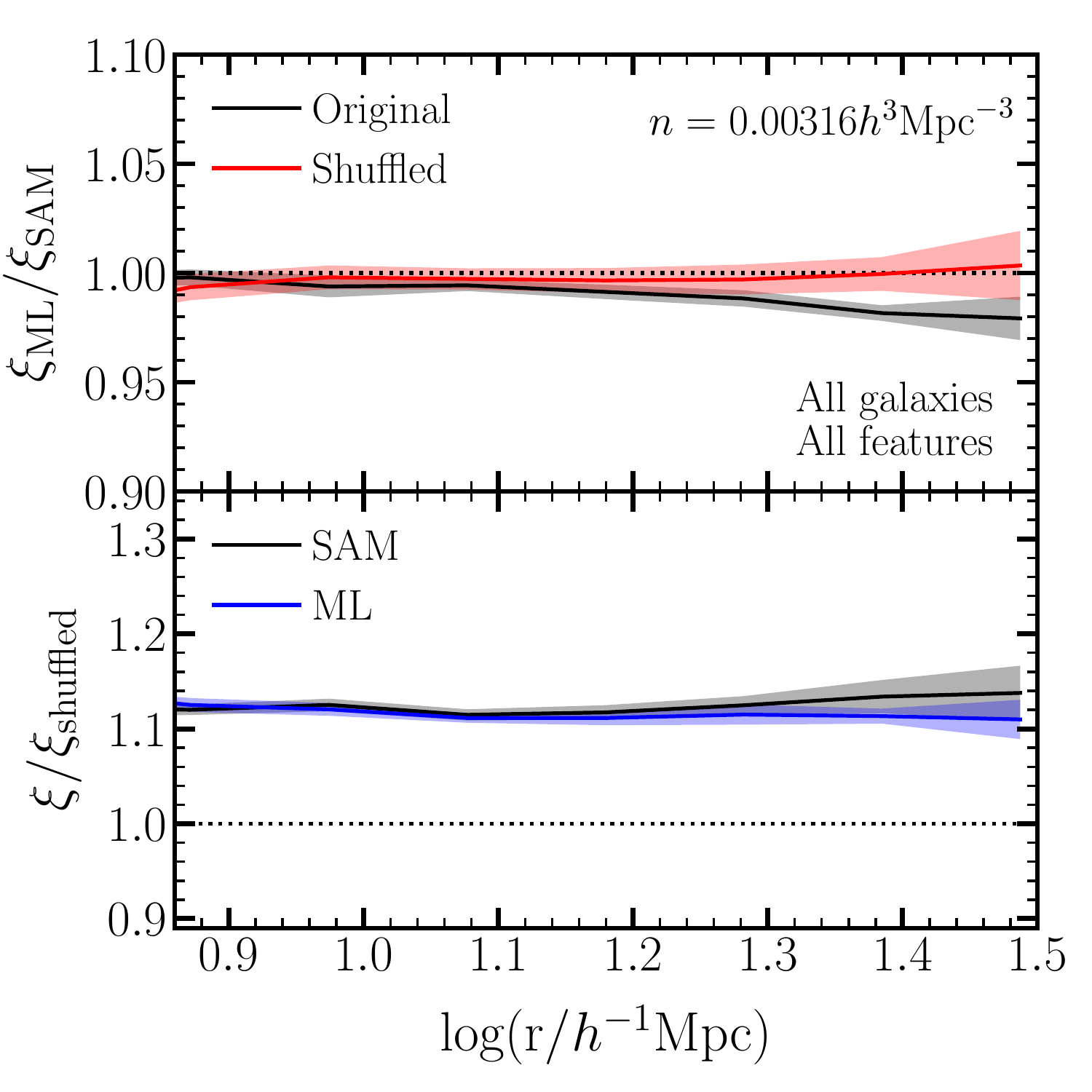}
	\end{subfigure}
	\hfill
\caption{
Similar to Figure~\ref{cluster_gab_all}, the ML predicted galaxy clustering and GAB with all features for the galaxy sample with number density $n=0.00316 \hmpcc$.
}
\label{xi_n1}
\end{figure*}

\begin{figure*}
	\centering
	\begin{subfigure}[h]{0.48\textwidth}
		\includegraphics[width=\textwidth]{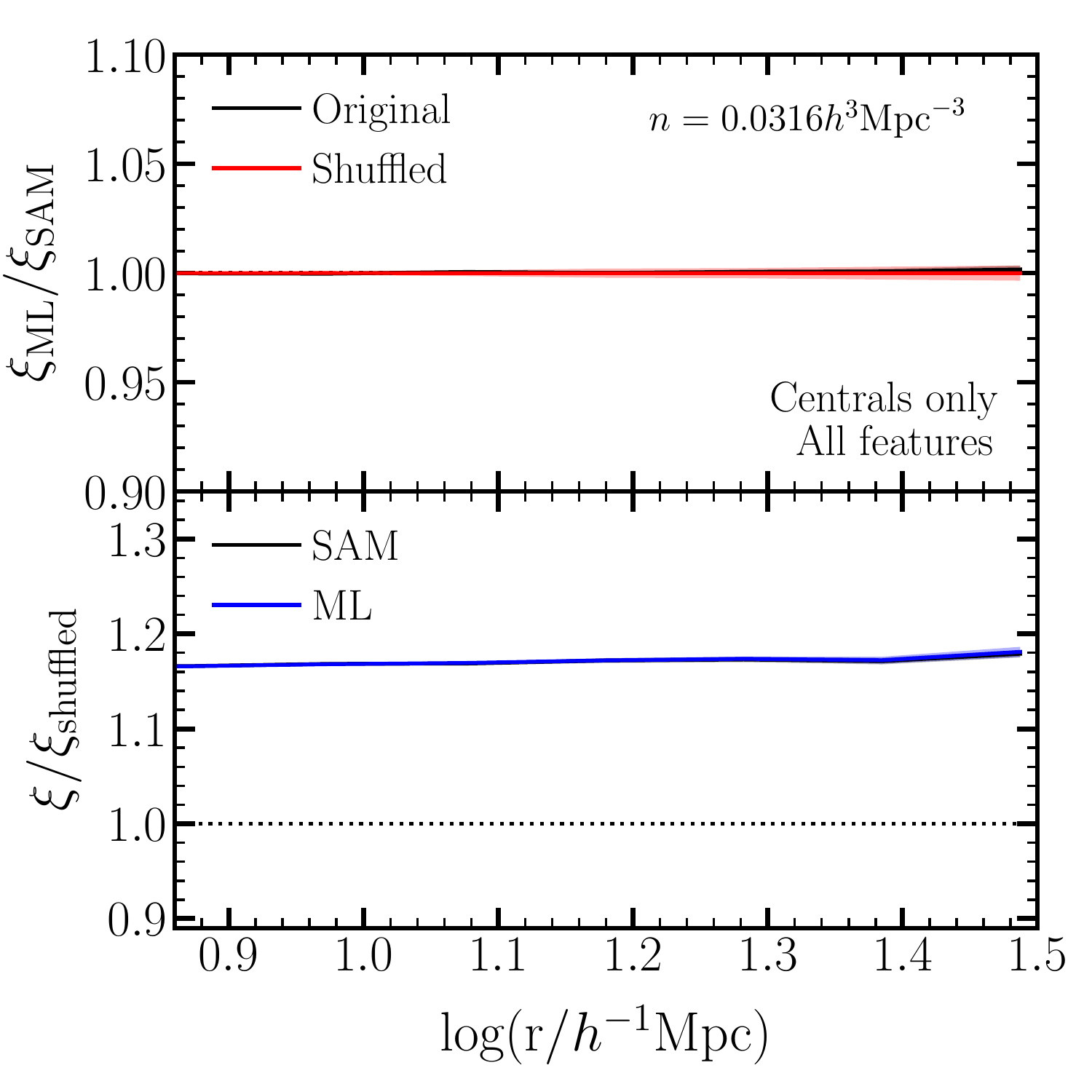}
	\end{subfigure}
	\hfill
	\begin{subfigure}[h]{0.48\textwidth}
        \includegraphics[width=\textwidth]{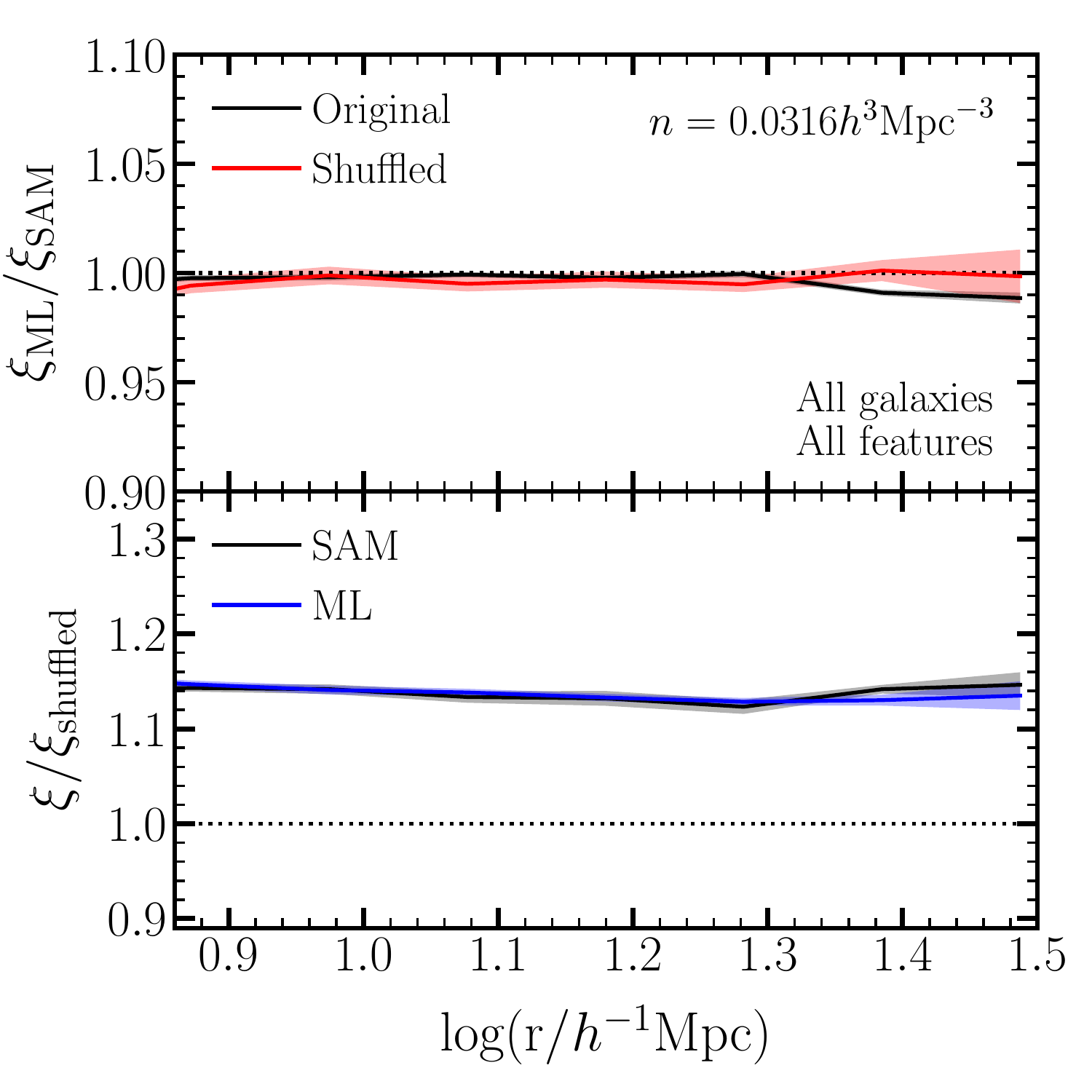}
	\end{subfigure}
	\hfill
\caption{
The same as in Figure~\ref{xi_n1} but for galaxy number density of $n=0.0316 \hmpcc$.
}
\label{xi_n3}
\end{figure*}

To further investigate the ability of RF models to reproduce the GAB, we perform a similar analysis to that presented in Section~\ref{allfeatures}, using all features available for the ML prediction, for two additional stellar-mass selected galaxy samples with $n=0.00316 \hmpcc$ and $n=0.0316 \hmpcc$. These correspond to stellar-mass thresholds of $3.88\times10^{10} \hmsun$ and $1.85 \times10^{9} \hmsun$, respectively.  The clustering results are shown in Figure~\ref{xi_n1} and Figure~\ref{xi_n3}, and are also included in Table~\ref{table:clustering}.

For the lowest number density sample, the results contain a higher level of noise, due to the smaller sample size. The $F_1$ and $R^2$ performance scores are correspondingly worse than for our default $n=0.01 \hmpcc$ sample, as well as the predicted clustering, especially on very large scales. This leads to a recovery of about 83\% of the GAB obtained for the central galaxies and 96\% recovery of the GAB for all (central and satellites) galaxies. However, as can be seen in Figure~\ref{xi_n1}, the larger uncertainties on these measurements imply a smaller level of discrepancy than a naive interpretation of these numbers. Furthermore, the SAM GAB measurements show an uncharacteristic scale-dependent behavior on the largest scales, which the ML predictions do not recover. This apparent scale dependence is likely just noise \citep{Xu2021}, such that the agreement is probably better than it seems.

On the other hand, for the sample with the highest number density, the sample size is larger accordingly, so that the measurement uncertainties and performance scores are better. The predicted clustering and GAB are all very close to 100\% in this case as expected.  We note that this sample includes also less massive galaxies resulting in slightly larger amount of GAB. We conclude that the accuracy of the ML predictions is fairly robust to the GAB level and not specific to the default $n=0.01 \hmpcc$ sample, but is somewhat sensitive to the level of noise as reflected by the size of the galaxy sample.

\section{Predicted occupancy variations}
\label{more_ov}

\begin{figure*}
	\centering
	\begin{subfigure}[h]{0.6\textwidth}
		\includegraphics[width=\textwidth]{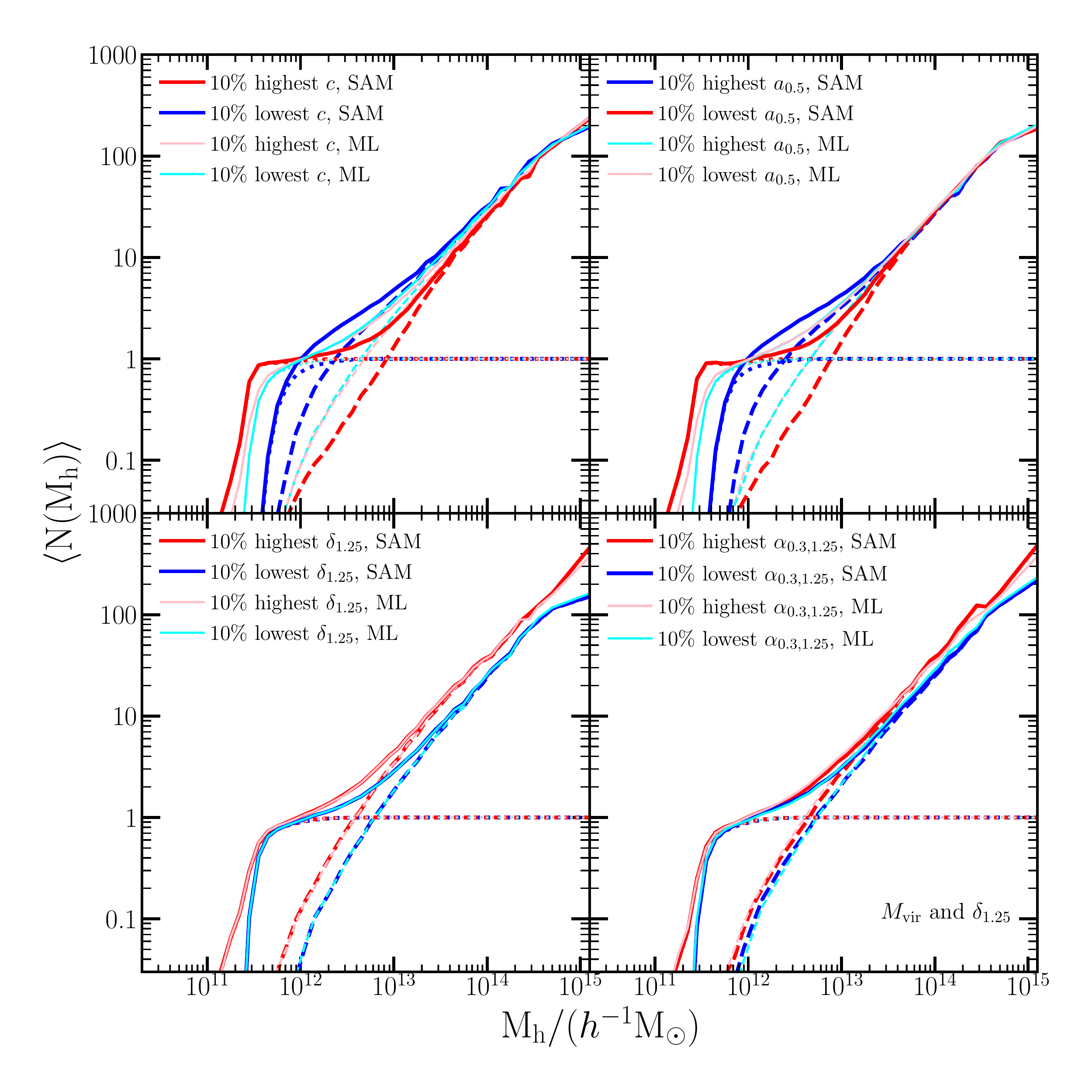}
	\end{subfigure}
	\hfill
\caption{Similar to Figure~\ref{ov_all} and Figure~\ref{ov_top}, the predicted OV with $c$, $a_{\rm 0.5}$, $\delta_{\rm 1.25}$, and $\alpha_{\rm 0.3,1.25}$, but now using only $M_{\rm vir}$ and $\delta_{\rm 1.25}$ as inputs for the RF algorithm for both centrals and satellites.
}
\label{ov_ME}
\end{figure*}  

\begin{figure*}
	\centering
	\begin{subfigure}[h]{0.6\textwidth}
		\includegraphics[width=\textwidth]{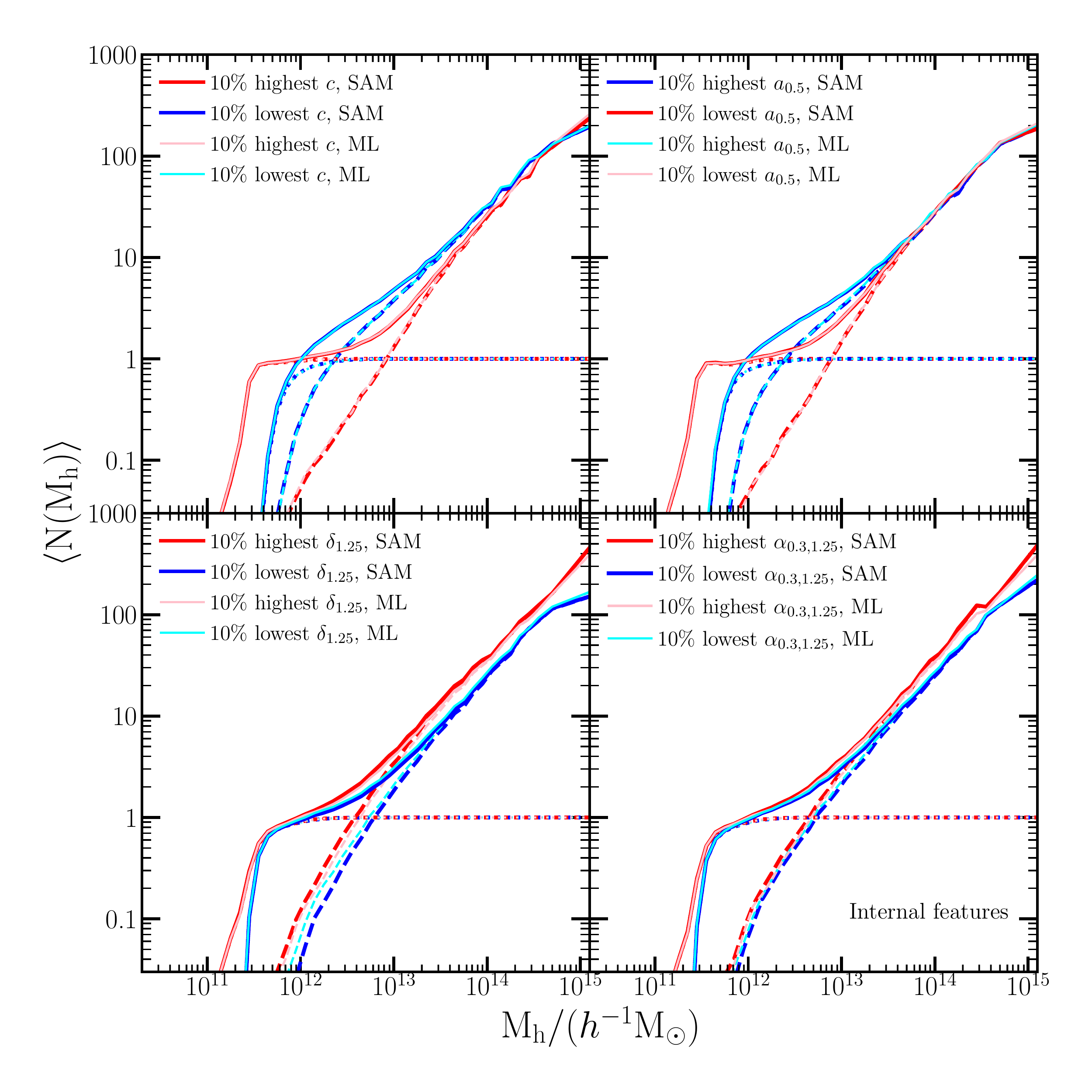}
	\end{subfigure}
	\hfill
\caption{Similar to Figure~\ref{ov_ME}, the predicted OV now with all internal halo properties as input features.
}
\label{ov_in}
\end{figure*}

\begin{figure*}
	\centering
	\begin{subfigure}[h]{0.6\textwidth}
		\includegraphics[width=\textwidth]{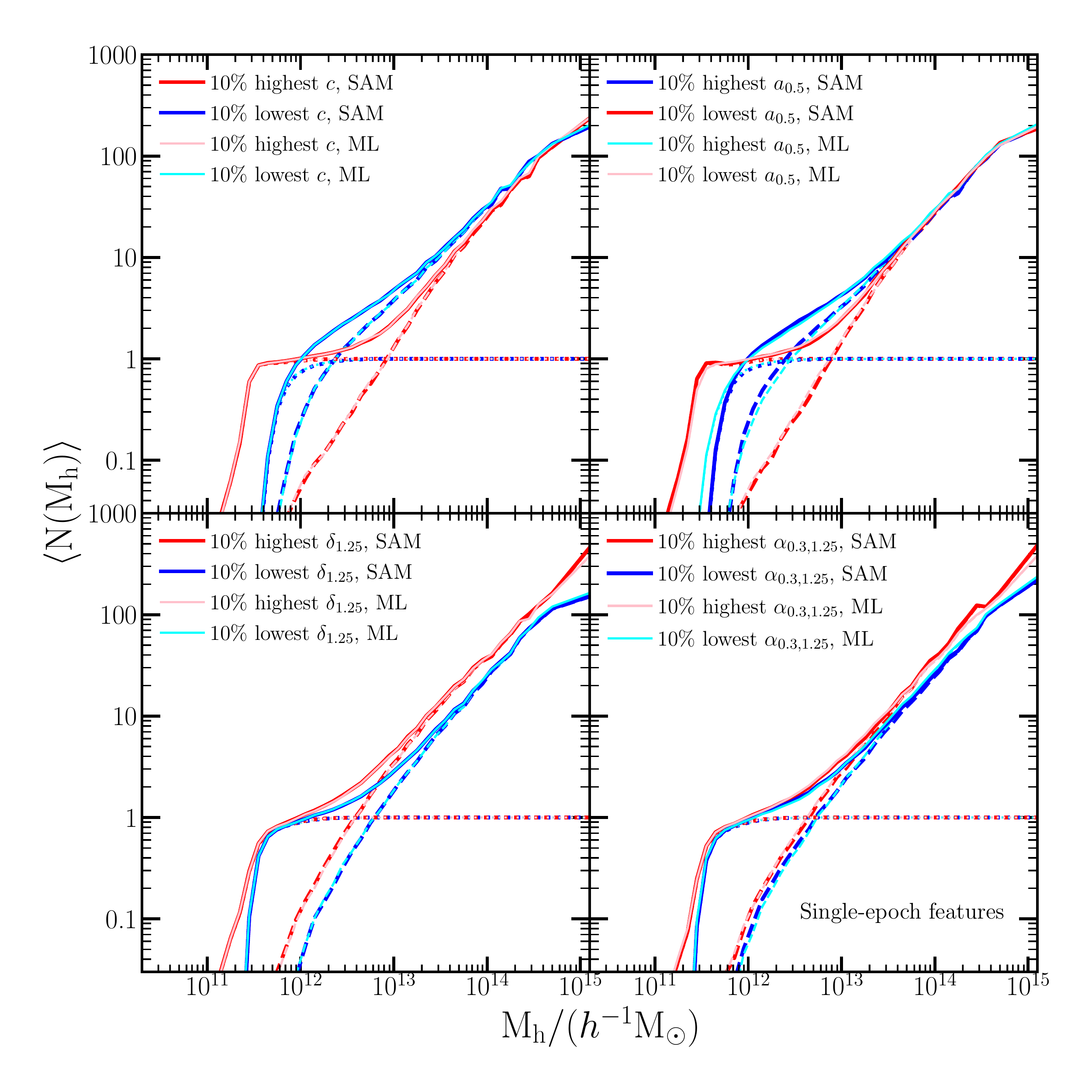}
	\end{subfigure}
	\hfill
\caption{Similar to Figure~\ref{ov_in}, the predicted OV here using only single-epoch properties.
}
\label{ov_non}
\end{figure*}

In this appendix, we provide the predicted OVs for the RF models in Section~\ref{mass_env} and Section~\ref{internal}. Figure~\ref{ov_ME} shows the predicted OVs by the RF models when using only halo mass and $\delta_{\rm 1.25}$ as input. Comparing to the SAM results, the OV with $\delta_{\rm 1.25}$ is accurately recovered as expected, as well the OV with $\alpha_{\rm 0.3,1.25}$ to a large extent since $\alpha_{\rm 0.3,1.25}$ is correlated with $\delta_{\rm 1.25}$. However, the predicted OV with either concentration or $a_{\rm 0.5}$ is not reproduced. For the centrals OV the trend with these properties is still there but to a much lesser degree than that of the SAM, indicated by the smaller difference of centrals occupations between upper and lower 10\% of the concentration and $a_{\rm 0.5}$. The satellites OV with these two internal properties is entirely missing, with identical satellite occupations for the upper and lower 10\% of the haloes. This may seem surprising initially since for the satellite galaxies $M_{\rm vir}$ and $\delta_{\rm 1.25}$ are two of the top four features, which are able to recover the OV and clustering well.  However, this arises due to the lack of any internal halo property other than mass as input for the RF models (while in the top features the halo concentration is included). Similar results for the OV were also obtained by \citet{Xu2021} when using $M_{\rm vir}$ and $\delta_{\rm 1.25}$. Despite the failure in recovering the OV with internal properties, the ML prediction based on mass and $\delta_{\rm 1.25}$ still reproduces the roughly correct level of clustering as in the SAM and a large fraction (0.92) of the GAB signal, as shown in Section~\ref{mass_env}.

Figure~\ref{ov_in} shows the predicted OVs of the RF models using all internal halo properties and no environmental measures (Section~\ref{internal}). In this case, we are able to reproduce quite well the OV dependences on all properties. Both the centrals and satellites OV with the internal properties $c$ and $a_{\rm 0.5}$ are recovered remarkably well, essentially by construction since these properties are included in the training. The centrals OV with the environmental properties $\alpha_{\rm 0.3,1.25}$ and $\delta_{\rm 1.25}$ are also well recovered. However, the predicted satellites OV with these properties is smaller than that of the SAM. These results are consistent with the feature importance discussed in Section~\ref{feature_imp}, where the environmental measures are among the top features for the satellite galaxies but not for the centrals. As a result, when excluding the environmental measures, the centrals-only clustering and GAB are fully recovered, but only 70\% of the GAB signal is reproduced when using both centrals and satellites. This indicates that the environment is important for reproducing the satellites OV and GAB, and can not be replaced with the impact of the internal halo properties. We conclude that with only internal properties, the centrals GAB can be well reproduced, but the environment is important for reproducing the full GAB of all galaxies.    

Finally, for completeness,  we also show here the OV for the single-epoch properties discussed in Section~\ref{nontree}. Since the concentration and $\delta_{\rm 1.25}$ are included in the input features for the RF models, their OVs are well recovered. The OV with $\alpha_{\rm 0.3,1.25}$ is also well reproduced, again likely due to the correlation with $\delta_{\rm 1.25}$.  The OV with age is accurately obtained for the satellites, but for the central galaxies it slightly deviates from the measurement in the SAM, producing a smaller OV. This may not be too surprising as $a_{\rm 0.5}$ is not included in the input features, since it requires the halo merger tree to be computed.  However, the concentration (which is included as a feature in this analysis) is correlated with $a_{\rm 0.5}$, and as such carries with it some (incomplete) information on age as well.  The resulting clustering and GAB are reasonably well reproduced.

\label{lastpage}

\end{document}